\documentclass[12pt]{article}

\usepackage{amsmath,amssymb,amsfonts,graphicx,amsfonts}
\usepackage[hidelinks]{hyperref}

\allowdisplaybreaks[4]

\date{}
\begin{document}
\begin{flushright}
\end{flushright}

\vspace{0.1cm}

\begin{center}
  {\LARGE
Quantum Lyapunov Spectrum
  }
\end{center}
\vspace{0.1cm}
\vspace{0.1cm}
\begin{center}

Hrant G{\sc haribyan}$^{a}$,
         Masanori H{\sc anada}$^{b}$,
         Brian S{\sc wingle}$^c$ and Masaki T{\sc ezuka}$^d$

\vspace{0.3cm}

$^a${\it Stanford Institute for Theoretical Physics,
Stanford University, Stanford, CA 94305, USA}

\vspace{0.2cm}
$^b$ {\it Department of Physics, University of Colorado, Boulder, Colorado 80309, USA}
\vspace{0.2cm}

$^c${\it
Condensed Matter Theory Center, Maryland Center for Fundamental Physics, Joint Center for
Quantum Information and Computer Science, and Department of Physics, University of Maryland,
College Park MD 20742, USA}\\

\vspace{0.2cm}
$^d${\it Department of Physics, Kyoto University, Kyoto 606-8502, Japan}\\

\end{center}

\vspace{1.5cm}

\begin{center}
  {\bf Abstract}
\end{center}

We introduce a simple quantum generalization of the spectrum of classical Lyapunov exponents. We apply it to the SYK and XXZ models, and study the Lyapunov growth and entropy production.
Our numerical results suggest that a black hole is not just the fastest scrambler, but also the fastest entropy generator.
We also study the statistical features of the quantum Lyapunov spectrum and
find universal random matrix behavior, which resembles the recently-found universality in classical chaos.
The random matrix behavior is lost when the system is deformed away from chaos, towards integrability or a many-body localized phase. We propose that quantum systems holographically dual to gravity satisfy this universality in a strong form. We further argue that the quantum Lyapunov spectrum contains important additional information beyond the largest Lyapunov exponent
and hence provides us with a better characterization of chaos in quantum systems.

\newpage
\tableofcontents
\section{Introduction}
\hspace{0.51cm}
Many-body quantum chaos is of fundamental interest in a variety of fields of physics,
including condensed  matter, quantum information, and quantum gravity. Considerable recent progress has come from the realization that it is possible, in some cases, to define a kind of quantum butterfly effect and a corresponding quantum Lyapunov exponent via so-called out-of-time-order correlators (OTOCs) \cite{larkin1969quasiclassical,Almheiri:2013hfa}. It was also discovered that the exponent so defined obeys a universal bound, the Maldacena-Shenker-Stanford (MSS) bound, $\lambda_N\le 2\pi T$ \cite{Maldacena:2015waa}, and that the bound is saturated by strongly coupled quantum systems holographically dual to Einstein gravity~\cite{Shenker2014,Kitaev_talk}.

This notion of quantum Lyapunov exponent has since received intense scrutiny; it is related to information scrambling~\cite{Shenker2014, Hayden2007, Sekino2008, Hosur2016} and thermalization~\cite{Deutsch1991, Srednicki1994,Tasaki1998a,Rigol2008}, it can be measured experimentally~\cite{swingle2016,Zhu2016,Yao2016a,Halpern2016,Halpern2017,Campisi2017,Yoshida2017,Garttner2016, Wei2016, Li2017a,  Meier2017}, it relates to operator growth~\cite{Nahum2017a, VonKeyserlingk2017, Xu2018, Roberts2018, Jonay2018, Mezei2018, You2018, Chen2018}, and can be computed either numerically or analytically in many model systems~\cite{Kitaev_talk, Nahum2017a, VonKeyserlingk2017, Xu2018, Sachdev1993, Gu2017, Luitz2017, Bohrdt2017a, Heyl2018, Lin2018, Nahum2017, Rakovszky2017, Khemani2017, Khemani2018,Lashkari2012, Shenker2014a,Aleiner2016, Swingle:2016jdj, Grozdanov2018,Patel2017,Swingle2018Resilience,Dressel2018SCweak,Menezes2018Sonic,Scaffidi2017}. However, to the best of our knowledge, these recent developments have focused almost exclusively on what, in retrospect, might be called the {\it largest} quantum Lyapunov exponent (two exceptions~\cite{hallam2018lyapunov,rozenbaum2018universal}, are discussed below). Yet at least for systems near a classical limit, the whole spectrum of Lyapunov exponents (the classical Lyapunov spectrum) makes sense and can contain additional useful information. For example, the sum of the positive Lyapunov exponents, called the Kolmogorov-Sinai (KS) entropy, characterizes the strength of chaos more precisely than the largest Lyapunov exponent alone.

Given these rapid developments, it is natural and desirable to attempt to extend the notion of a classical Lyapunov spectrum to general quantum systems away from the classical limit. In this paper, we give one definition of a quantum Lyapunov spectrum which makes sense for any many-body quantum system. There are several motivations for this study. One question is do black holes become more or less chaotic as they grow in size? To answer this basic question, we need to define the strength of chaos precisely.
As explained in Sec.~\ref{sec:classical_chaos}, the KS entropy is a better indicator than the largest exponent alone.
It is likely that, as a black hole grows, the KS entropy increases, while the largest Lyapunov exponent decreases \cite{Berkowitz:2016znt}.
Hence looking at the entire Lyapunov spectrum gives a different picture of chaos in black holes than the largest Lyapunov exponent alone.
Another motivation is universality in the classical Lyapunov spectrum \cite{Hanada:2017xrv}:
in some chaotic systems, the classical spectrum converges to a random matrix theory (RMT) spectrum, with the time scale for the onset of
universality seemingly related to the strength of chaos. In particular, a matrix model of black holes \cite{Banks:1996vh,deWit:1988wri,Itzhaki:1998dd} shows the universality from $t=0$, indicating a possible signature of gravity in the Lyapunov spectrum.

In this paper, we introduce a generalization of the classical Lyapunov spectrum to quantum many-body systems,
including systems far from any classical limit. Our definition has a natural physical interpretation and, when the classical limit can be taken, it reduces to the usual classical Lyapunov spectrum. Furthermore, the largest Lyapunov exponent agrees with the usual quantum Lyapunov exponent in the literature as obtained from OTOCs at sufficiently late times.

To elucidate the physics of our definition, we study two systems, the non-local Sachdev-Ye-Kitaev (SYK) model \cite{Kitaev_talk,Maldacena:2016hyu,Sachdev:2015efa} and the local XXZ model with random magnetic field (see e.g.~\cite{PhysRevB.91.081103}). In the case of SYK, disordered couplings are part of the basic definition. In the case of XXZ, the model is considered in the isotropic (Heisenberg) limit with an additional disordered magnetic field. We analyze both by performing systematic exact diagonalization studies of the quantum Lyapunov spectrum including disorder averaging. Our main results can be summarized as follows.
\begin{itemize}
\item
We observe a period of approximately exponential growth, meaning an approximately time-independent Lyapunov spectrum, for the nonlocal SYK model. For the local XXZ model we observe a brief period of approximately exponential growth followed by a period of power law growth.

\item
We numerically demonstrate that a naive generalization of the classical KS entropy --- just the sum of positive quantum Lyapunov exponents ---
is close to the production rate of the entanglement entropy, with a proper normalization needed for the connection to classical coarse-grained entropy.

\item
We also suggest that this `quantum KS entropy' is maximized when the quantum system has dual gravity description on the black hole background. In other words, we suggest that black hole is not just the fastest scrambler but also the fastest entropy generator.

\item
For both SYK and XXZ models in the appropriate regime we observe random matrix theory (RMT) behavior for the quantum Lyapunov spectrum. This universality fails in expected ways, including for integrable and localized systems.

\end{itemize}

The outline of the remainder of the paper is as follows. Section 2 recalls the basic definition of the classical Lyapunov spectrum. Section 3 introduces the SYK and XXZ models. Section 4 defines the notion of a quantum Lyapunov exponent. Section 5 studies the growth characteristics of the quantum spectrum with time. Section 6 studies the distribution of the quantum Lyapunov spectrum, establishing a link to random matrix statistics. Section 7 contains concluding remarks and outlook.

\section{Lyapunov Spectrum in Classical Chaos}\label{sec:classical_chaos}
\hspace{0.51cm}
First let us see how the classical Lyapunov exponents are defined. For simplicity we consider only Hamiltonian systems.
Suppose the phase space is described by coordinate variables $x_i\ (i=1,2,\cdots,N)$ and conjugate momenta $p_i(i=1,2,\cdots,N)$.
We use $z_i\ (i=1,2,\cdots,2N)$ to denote $x_i$ and $p_i$ together.
The sensitivity to the initial condition at $t=0$ is captured by
\begin{eqnarray}
M_{ij}(t)\equiv\frac{\delta z_i(t)}{\delta z_j(0)}.
\label{Mij-classical}
\end{eqnarray}
The finite-time Lyapunov exponents $\lambda_i(t)$ are defined by
\begin{eqnarray}
\lambda_i(t)\equiv\frac{1}{t}\log s_i(t),
\end{eqnarray}
where $s_i(t)$ are singular values of $M_{ij}(t)$.
These exponents converge to constant values at sufficiently late time.
Note that they are calculated for each initial condition, which is analogous to the microstate in the quantum theory.
Often the average over initial conditions is taken.
Note also that we can calculate the exponents from the eigenvalues of
\begin{eqnarray}
L_{ij}(t)=[M^\dagger(t) M(t)]_{ij}=M^\ast_{ki}(t) M_{kj}(t),
\end{eqnarray}
which are $\left(s_i(t)\right)^2$.

Often the limit $t\to\infty$ is considered. In this paper, we will be interested in the finite-$t$ behavior.
\subsection{Kolmogorov-Sinai Entropy}\label{sec:KSentropy}
\begin{figure}[htbp]
\begin{center}
\scalebox{0.6}{
\includegraphics{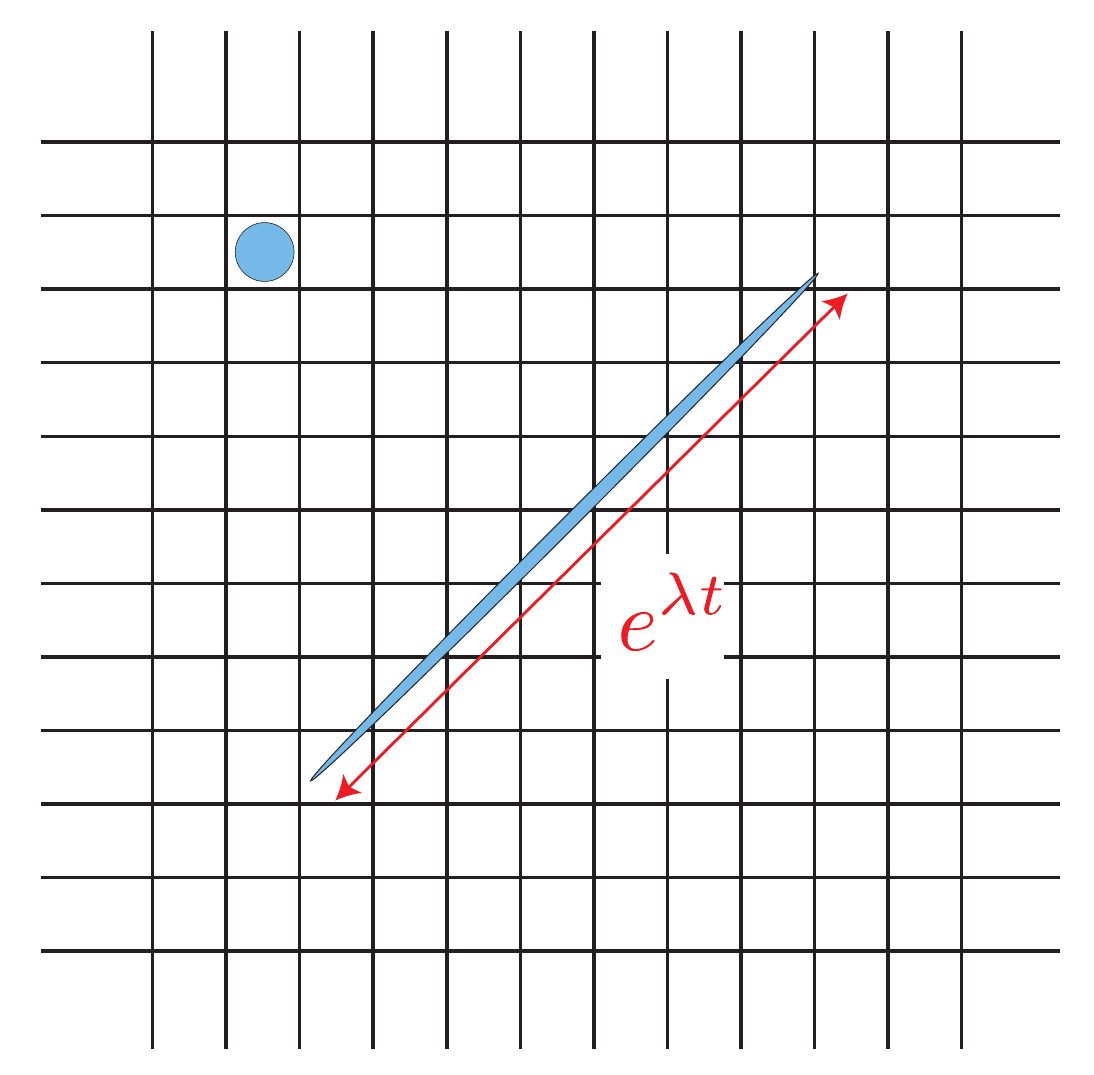}}
\end{center}
\caption{Suppose we know that the initial condition is contained in the blue disk in the upper left corner.
As this region evolves with time, although the volume is conserved due to the Liouville theorem,
the shape changes nontrivially, in particular, it is stretched exponentially in several directions.
The rate is governed by the positive Lyapunov exponents.
}\label{fig:coarse-graining-KS}
\end{figure}

\hspace{0.51cm}
Suppose our knowledge about the initial condition is limited and we only know that it is in a small region in the phase space,
say the blue disk in Fig.~\ref{fig:coarse-graining-KS}. As time passes by, this region is stretched along some directions
and compressed along the others.
If we introduce a grid like in Fig.~\ref{fig:coarse-graining-KS} and count the number of cells
needed for covering the region, more and more cells are needed at later time;
the number scales as
\begin{eqnarray}
\prod_{\lambda>0}e^{\lambda t} = e^{h_{\rm KS}t},
\qquad
h_{\rm KS} \equiv \sum_{\lambda>0}\lambda.
\end{eqnarray}
This exponential growth characterizes the loss of our knowledge about a given initial condition.
The coarse-grained entropy, which is the log of the uncertainty, increases as $h_{\rm KS}t$.
The Kolmogorov-Sinai (KS) entropy $h_{\rm KS}$ is the growth rate of the coarse-grained entropy~\cite{PhysRevLett.82.520} (although for exceptions see Ref.~\cite{PhysRevE.71.016118}).~\footnote{Strictly speaking, $d(h_\mathrm{KS}t)/dt$ is the entropy production rate.
At late time, $h_\mathrm{KS}$ converges to a constant, and hence $d(h_\mathrm{KS}t)/dt=h_\mathrm{KS}$.}

The quantum counterpart of the coarse-grained entropy is the entanglement entropy.
Indeed, the growth of the entanglement entropy of a subregion of a system is an essential part of quantum thermalization.
It would be desirable if the KS entropy $h_{\rm KS}$ could be identified with the growth rate of the entanglement entropy as well, at least near the classical limit.

\subsubsection{Application to Black Hole}\label{sec:black-hole-merger}
\begin{figure}[htbp]
\begin{center}
\scalebox{0.5}{
\includegraphics{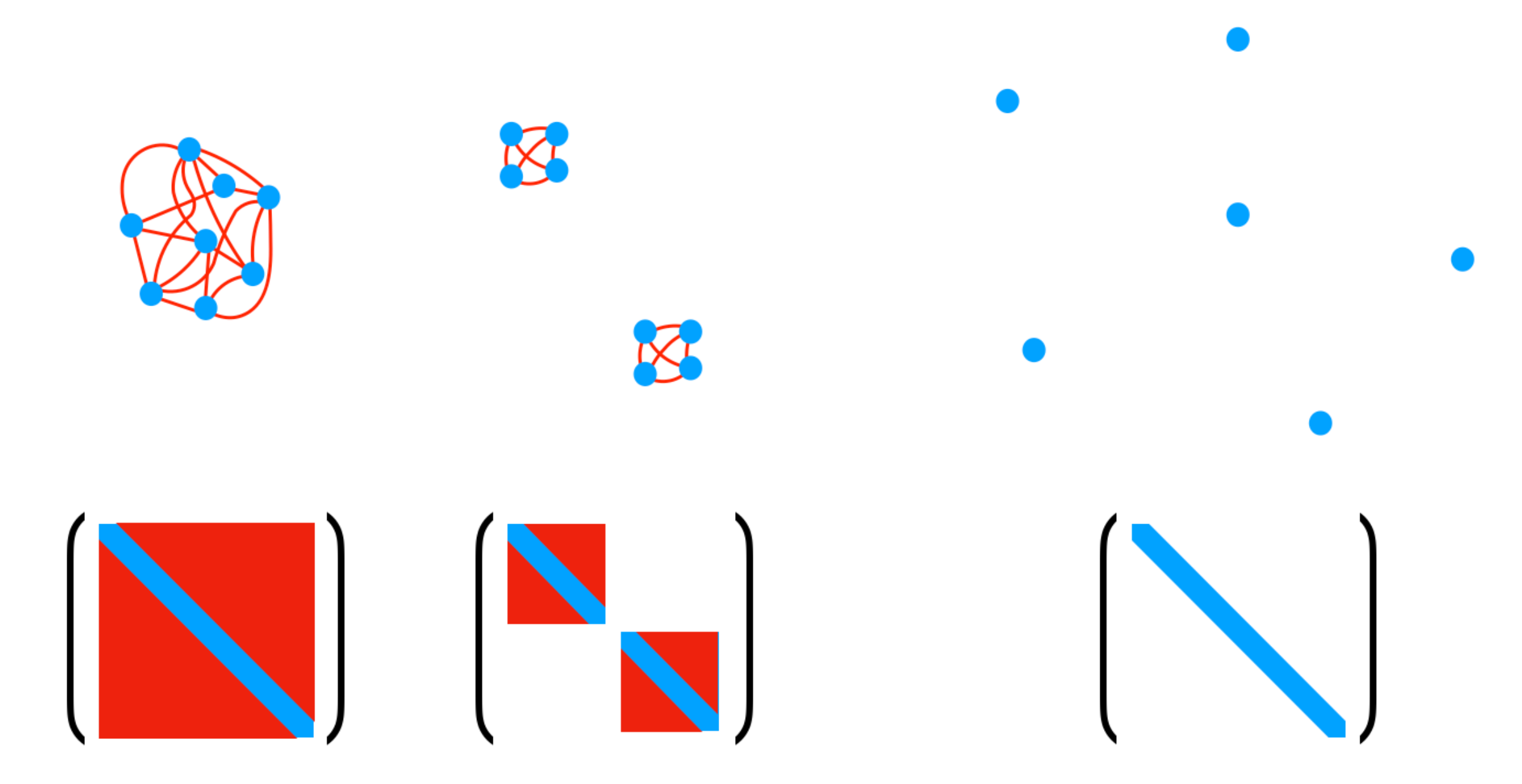}}
\end{center}
\caption{Matrix configurations for one big black hole (left), two black holes (middle) and the gas of D0-branes (right).
}\label{fig:matrix-black-holes}
\end{figure}

\hspace{0.51cm}
In the sense discussed above, the spreading of information in phase space is better characterized by the KS entropy than the largest Lyapunov exponent alone.
This is particularly so for a black hole.
In order to understand why, let us consider the matrix model of black hole
made of D0-branes and strings \cite{Banks:1996vh,deWit:1988wri,Itzhaki:1998dd},
which is the original setup used by Sekino and Susskind to argue for fast scrambling \cite{Sekino2008}. (Essentially the same argument applies to black hole described by other gauge theories as well.)
This model contains nine $N\times N$ Hermitian matrices as bosonic degrees of freedom.
Diagonal and off-diagonal elements are interpreted as the locations of D0-branes in nine-dimensional space
and open strings connecting D0-branes. As shown in Fig.~\ref{fig:matrix-black-holes},
a big black hole consisting of all $N$ D0-branes and many open string excitations is described by fully excited
$N\times N$ matrices, while block diagonal configurations describe several black holes
and the diagonal matrices describe a gas of D0-branes without strings.

Suppose two black holes consisting of $\frac{N}{2}\times\frac{N}{2}$ blocks (the middle of Fig.~\ref{fig:matrix-black-holes}) come close and
merge to a single black hole (the left of Fig.~\ref{fig:matrix-black-holes}).
During this process, open strings stretched between two black holes (off-diagonal blocks) become shorter,
and hence lighter, and eventually get excited. In this way the number of dynamical degrees of freedom are doubled,
and hence temperature, which is roughly equivalent to the energy per degree of freedom,
goes down \cite{Berkowitz:2016znt}.
For example, in the highly stringy region where the matrix model admits a classical description,
the energy is $E=6N^2T_{\rm fin}=2\times 6\left(\frac{N}{2}\right)^2T_{\rm init}$, and hence $T_{\rm fin}=\frac{1}{2}T_{\rm init}$.
This is the gauge theory description of a key property of a black hole discovered in Hawking's seminal paper \cite{Hawking:1974sw}:
a black hole in flat spacetime cools down as it grows in size.
As the black hole cools down, the largest Lyapunov exponent becomes smaller; in the highly stringy region,
it scales as $T^{1/4}$ \cite{Gur-Ari:2015rcq}.
Hence the largest Lyapunov exponent becomes smaller.
However the number of positive Lyapunov exponents increases because of the dynamical increase of
the degrees of freedom. Hence there are two competing contributions --- the decrease of the temperature pushes down the KS entropy, while the increase of the degrees of freedom pushes it up.
A careful evaluation of these effects shows that the KS entropy actually increases as black hole grows \cite{Berkowitz:2016znt}.
Hence a bigger black hole is a faster entropy generator.

\subsection{Universality in the Lyapunov Spectrum} \label{sec:universality_classical}
\hspace{0.51cm}
The Lyapunov spectrum of the classical limit of the matrix model of black holes has been studied in Ref.~\cite{Gur-Ari:2015rcq} and Ref.~\cite{Hanada:2017xrv}.
In particular, the level correlations of the Lyapunov spectrum have been studied. Ref.~\cite{Hanada:2017xrv} studied other classically chaotic systems as well, and found a universal random-matrix description:
{\it when the number of degrees of freedom is sufficiently large, the level correlation of the finite-time Lyapunov exponents converges to the one determined by a Gaussian random matrix theory}.
Furthermore, {\it the matrix model of a black hole shows this universality already at $t=0$}.

If this universality can be generalized to quantum chaos, it should serve as a new characterization of quantum chaos.
Note that this universality is different from (though it might be related to) the usual Wigner-Dyson universality
of the energy spectrum.
First of all we are looking at different things --- Lyapunov exponents and energy levels ---
and furthermore the onsets of the universal distribution are observed at different time scales:
while the latter sets in at rather late time (see \cite{Gharibyan:2018jrp} for a detailed discussion),
the former can set in at earlier time, as it is there already at $t=0$ for the matrix model of black holes.
As we will see later in this paper, the SYK model shows this Lyapunov universality at $t=0$ as well.
\subsection{Technical Remarks}
\hspace{0.51cm}
A few technical remarks are in order here. Firstly, there is an ambiguity in the definition of the Lyapunov exponents at finite time associated with a choice of variables,
although there exists a unique $t\to\infty$ limit. If we choose another basis $z'=z'(z)$, $M_{ij}(t)=\frac{\delta z_i(t)}{\delta z_j(0)}$ changes to
\begin{eqnarray}
M'_{ij}(t)
=
\frac{\delta z'_i(t)}{\delta z'_j(0)}
=
\frac{\delta z'_i(t)}{\delta z_k(t)}
\frac{\delta z_k(t)}{\delta z_l(0)}
\frac{\delta z_l(0)}{\delta z'_j(0)}
=
J_{ik}M_{kl}(t)J^{-1}_{lj},
\end{eqnarray}
where $J_{ij}=\frac{\delta z'_i}{\delta z_j}$ is the Jacobian matrix.
Still, this definition captures the Lyapunov growth.
Furthermore, at sufficiently late time, the Lyapunov exponents converge to the same values.
The choice of the basis may affect the details at early time, and hence we will choose a natural one
which makes the physical interpretation more transparent.

Secondly, for the Hamiltonian system, $M$ is symplectic, and the Lyapunov exponents form pairs of $+\lambda$ and $-\lambda$.
It provides us with an easy way to estimate the numerical error, namely this pair structure is gone when the error accumulates too much.
This property does not necessarily hold for the quantum Lyapunov exponents introduced in Sec.~\ref{sec:quantum_Lyapunov_definition}.

Thirdly, the finite-time Lyapunov exponents depend on the initial condition $z_i(0)$.
In the thermodynamic limit, as long as a sufficiently generic initial condition is chosen,
the same global structure (overall distribution) and microscopic spectral properties are obtained.
Practically, because our simulation is always with finitely many degrees of freedom,
we consider the average over many samples.
See Refs.~\cite{Hanada:2017xrv,Gur-Ari:2015rcq} for details.

\section{Models}\label{sec:models}
\hspace{0.51cm}
We will study two physically distinct models. The first will be Sachdev-Ye-Kitaev (SYK) model of four-local Majorana fermions coupled randomly and non-locally. The second model is the XXZ spin chain composed of nearest neighbor spin interactions and random magnetic field along the $z$-axis.

\subsection{Sachdev-Ye-Kitaev (SYK)}\label{sec:SYK_definition}
\hspace{0.51cm}
The SYK model \cite{Kitaev_talk,Maldacena:2016hyu,Sachdev:2015efa} (see Ref.~\cite{Rosenhaus:2018dtp} for a recent review) with $N$ Majorana fermions is given by
\begin{eqnarray}
\hat{H}=\sqrt{\frac{6}{N^3}}\sum_{i<j<k<l}J_{ijkl}\hat{\psi}_i\hat{\psi}_j\hat{\psi}_k\hat{\psi}_l, \label{Hamiltonian_SYK}
\end{eqnarray}
where the anti-commutation relations are given by
\begin{eqnarray}
\{\hat{\psi}_i,\hat{\psi}_j\}=\delta_{ij}
\end{eqnarray}
and $J_{ijkl}$ is random Gaussian with mean zero and standard deviation $J$.
We set $J$ to be 1, so all times are measured in units of $1/J$. The dimension of the Hilbert space is $2^{N/2}$.
We will also consider slightly more general version
\begin{eqnarray}
\hat{H}
=
\sqrt{\frac{6}{N^3}}\sum_{i<j<k<l}J_{ijkl}\hat{\psi}_i\hat{\psi}_j\hat{\psi}_k\hat{\psi}_l
+
\frac{\sqrt{-1}}{\sqrt{N}}\sum_{i<j}K_{ij}\hat{\psi}_i\hat{\psi}_j,
\label{eqn:q=2deformedSYK}
\end{eqnarray}
where $K_{ij}$ is also a random Gaussian variable with mean zero and standard deviation $K$.
At $K=0$, this model is `maximally chaotic' at low temperatures, in the sense that the MSS bound \cite{Kitaev_talk,Maldacena:2016hyu} is saturated.
When $K>0$, it is not chaotic at sufficiently low temperature,
while it remains chaotic at high temperature \cite{Garcia-Garcia:2017bkg,Nosaka:2018iat}.

\subsection{XXZ Spin Chain}\label{sec:XXZ_definition}
\hspace{0.51cm}
We will also consider one-dimensional spin chain
\begin{eqnarray}
\hat{H}
=
\sum_{i=1}^{N_\mathrm{site}}\left(
\frac{1}{4}
\vec{\sigma}_i\vec{\sigma}_{i+1}
+
\frac{w_{i}}{2}\sigma_{z,i}
\right),
\label{eqn:XXZ}
\end{eqnarray}
with periodic boundary condition $\sigma_{N_{\rm site}+1}=\sigma_1$.
$w_i$ is the random magnetic field along $z$-direction, chosen to be uniform random number between $[-W,+W]$.
At $W\gtrsim 2.75$, most of the energy eigenstates are in the many-body localized (MBL) phase \cite{PhysRevB.91.081103,serbyn1507criterion}.
(For the physics of the MBL phase, see e.g.~\cite{PhysRev.109.1492,PhysRevLett.95.206603,BASKO20061126,aleiner:hal-00543657}; for OTOC calculations see \cite{Swingle2016a, Chen2016a, Fan2017, Huang2017,Sahu2018}.) As $W$ is lowered, ergodic phase expands from the center of the spectrum, and gradually
the system becomes ergodic except for a small region at low and high energy regions. The boundary between the ergodic and MBL phases can be obscure when the system size is small.

\section{A Definition: Quantum Lyapunov Exponents}\label{sec:quantum_Lyapunov_definition}
\hspace{0.51cm}

A natural quantum analogue of $M_{ij}(t)$ defined by Eq.~\eqref{Mij-classical} is\footnote{
Here we assumed a bosonic system. We will consider a fermionic system (SYK model) later.
}
\begin{eqnarray}
\hat{M}_{ij}(t)\equiv\sqrt{-1}[\hat{z}_i(t),\hat{\Pi}_{j}(0)],
\label{def-Mij}
\end{eqnarray}
where $\hat{\Pi}_{j}$ is the canonical conjugate of $\hat{z}_j$.
For a given state $|\phi\rangle$, $\hat{M}_{ij}(t)|\phi\rangle$ can grow exponentially,
and
\begin{eqnarray}
L_{ij}^{(\phi)}(t)
\equiv
\langle\phi|\hat{M}^\ast_{ki}(t)\hat{M}_{kj}(t)|\phi\rangle
\label{def-Lij}
\end{eqnarray}
(where $^\ast$ is a conjugate as an operator acting on the Hilbert space)
is a natural counterpart of $L_{ij}(t)$ in the classical theory.\footnote{
Note that $\langle\phi|\hat{M}_{ij}(t)|\phi\rangle$ cannot capture the growth properly, because
the overlap between $\hat{M}_{ij}(t)|\phi\rangle$ and $|\phi\rangle$ becomes exponentially small.
}
From this we can define the Lyapunov exponents $\lambda_i^{(\phi)}(t)$.

The classical Lyapunov exponents in the Hamiltonian systems have degeneracy $\pm\lambda$.
The quantum Lyapunov exponents defined above do not necessarily have such degeneracy.

Below we omit $(\phi)$ in $\lambda_i^{(\phi)}(t)$, because we do not think there is a risk of confusion.
We will take $|\phi\rangle$ to be energy eigenstates.

\subsection{Lyapunov Exponents in SYK}
\hspace{0.51cm}
As a natural counterpart of Eq.~\eqref{def-Mij} in a fermionic system,
we can use
\begin{eqnarray}
\hat{M}_{ij}(t)=\{\hat{\psi}_i(t),\hat{\psi}_j(0)\}.
\end{eqnarray}

We will again take $|\phi\rangle$ in Eq.~\eqref{def-Lij} to be energy eigenstates. When $K$ is zero,
for $N$ not a multiple of eight, the eigenstates of the Hamiltonian are doubly degenerate due to a symmetry.
If $N$ is not a multiple of four, the degeneracy occurs within each parity sector.
In order to avoid this uncertainty, we will not consider $K=0$, instead we consider very small but finite value of $K$. For each $|\phi\rangle$, we obtain $N$ Lyapunov exponents, $\lambda_1\le\lambda_2\le\cdots\le\lambda_N$.

\subsection{Lyapunov Exponents in XXZ}
\hspace{0.51cm}
Some caution is in order when we define the Lyapunov exponents for the XXZ spin chain. Because the total $z$-spin $S^{\rm (total)}_z=\frac{1}{2}\sum_i\sigma_{z,i}$ commutes with the Hamiltonian, it is better to define $\hat{M}$ so that it commutes with $S^{\rm (total)}_z$ as well,
in order to avoid a mixture of different spin sectors which may complicate the analysis.
Furthermore $\sigma_x$, $\sigma_y$ and $\sigma_z$ are redundant, in the sense that
$\sigma_x\sigma_y=\sqrt{-1}\sigma_z$.

One possible option is to use the fermion representation obtained by the Jordan-Wigner transformation,
\begin{eqnarray}
\hat{\psi}_1
&=&
\left(\sigma_1\otimes\textbf{1}_2\otimes\cdots\otimes\textbf{1}_2\right)/\sqrt{2},
\nonumber\\
\hat{\psi}_2
&=&
\left(\sigma_2\otimes\textbf{1}_2\otimes\cdots\otimes\textbf{1}_2\right)/\sqrt{2},
\nonumber\\
\hat{\psi}_3
&=&
\left(\sigma_3\otimes\sigma_1\otimes\cdots\otimes\textbf{1}_2\right)/\sqrt{2},
\nonumber\\
\hat{\psi}_4
&=&
\left(\sigma_3\otimes\sigma_2\otimes\cdots\otimes\textbf{1}_2\right)/\sqrt{2},
\nonumber\\
& &\qquad\cdots
\nonumber\\
\hat{\psi}_{2N_{site}-1}
&=&
\left(\sigma_3\otimes\sigma_3\otimes\cdots\otimes\sigma_1\right)/\sqrt{2},
\nonumber\\
\hat{\psi}_{2N_{site}}
&=&
\left(\sigma_3\otimes\sigma_3\otimes\cdots\otimes\sigma_2\right)/\sqrt{2},
\end{eqnarray}
which satisfy the standard anticommutation relation $\{\hat{\psi}_i,\hat{\psi}_j\}=\delta_{ij}$.

Still, it is probably better if $\hat{M}$ is compatible with the locality which is manifest in terms of the $\vec{\sigma}_i$ variables.
Then there are several other options such as
\begin{eqnarray}
\sqrt{-1}[\sigma_{z,i}(t),\sigma_{z,j}(0)]
\end{eqnarray}
and
\begin{eqnarray}
[\sigma_{+,i}(t),\sigma_{-,j}(0)],
\end{eqnarray}
where $\sigma_{\pm}=\frac{\sigma_x+i\sigma_y}{2}$. Note that the latter is neither Hermitian or skew-Hermitian.
The former vanishes at $t=0$, which makes it difficult to define the Lyapunov growth precisely at early time,
while the latter gives $[\sigma_{+,i}(0),\sigma_{-,j}(0)]=\sigma_z\delta_{ij}$.
Here we use the latter:
\begin{eqnarray}
\hat{M}_{ij}\equiv[\sigma_{+,i}(t),\sigma_{-,j}(0)],
\qquad
L_{ij}^{(\phi)}(t)
\equiv
\langle\phi|\hat{M}^\ast_{ki}(t)\hat{M}_{kj}(t)|\phi\rangle.
\label{transfer_matrix_XXZ}
\end{eqnarray}
Then $L_{ij}^{(\phi)}(0)=\delta_{ij}$. The physical interpretation is clear: $\sigma_{+,i}(t)$ and $\sigma_{-,j}(0)$ creates/annihilates
the $z$-spin at point $i, j$ and time $t,0$, respectively.
\subsection{Kolmogorov-Sinai Entropy}
\hspace{0.51cm}

For classical systems, the Kolmogorov-Sinai entropy $h_\mathrm{KS}$ is defined as the sum of all positive Lyapunov exponents.
We use the same definition here, by using the Lyapunov exponents we have defined in this paper:
\begin{eqnarray}
h_{\rm KS} = \sum_{\lambda_i>0}\lambda_i.\label{KS-entropy-def}
\end{eqnarray}
Away from the classical limit, the properties of this quantity are not immediately clear.
It will be studied in Sec.~\ref{sec:SYK-EE-vs-KS}.

This `KS entropy' is not necessarily the same as the definition in other literature.
See Refs.~\cite{alicki2002information,alicki1998quantum,mendes1995entropy,man2000lyapunov,kunihiro2009towards,Cotler:2017anu} for other approaches.
\subsection{Is the `perturbation' actually small?}\label{sec:perturbation_size}
\hspace{0.51cm}
In the study of quantum chaos based on OTOCs, and also in our approach, one usually assumes that the perturbation
--- multiplication of local operators $\hat{V}$ and $\hat{W}$ --- does not change the state too much.
This should be the case when the system size is sufficiently large provided $W$ and $V$ are few-body operators. However it is not necessarily the case in actual numerical calculations.
Below we see that, with realistic system sizes within our reach, this assumption is not valid  for the SYK model and hence
some care is needed when we extract physics from the numerical data.
\subsubsection*{SYK}
\begin{figure}[htbp]
\begin{center}
\rotatebox{-90}{
\includegraphics[width=5.cm]{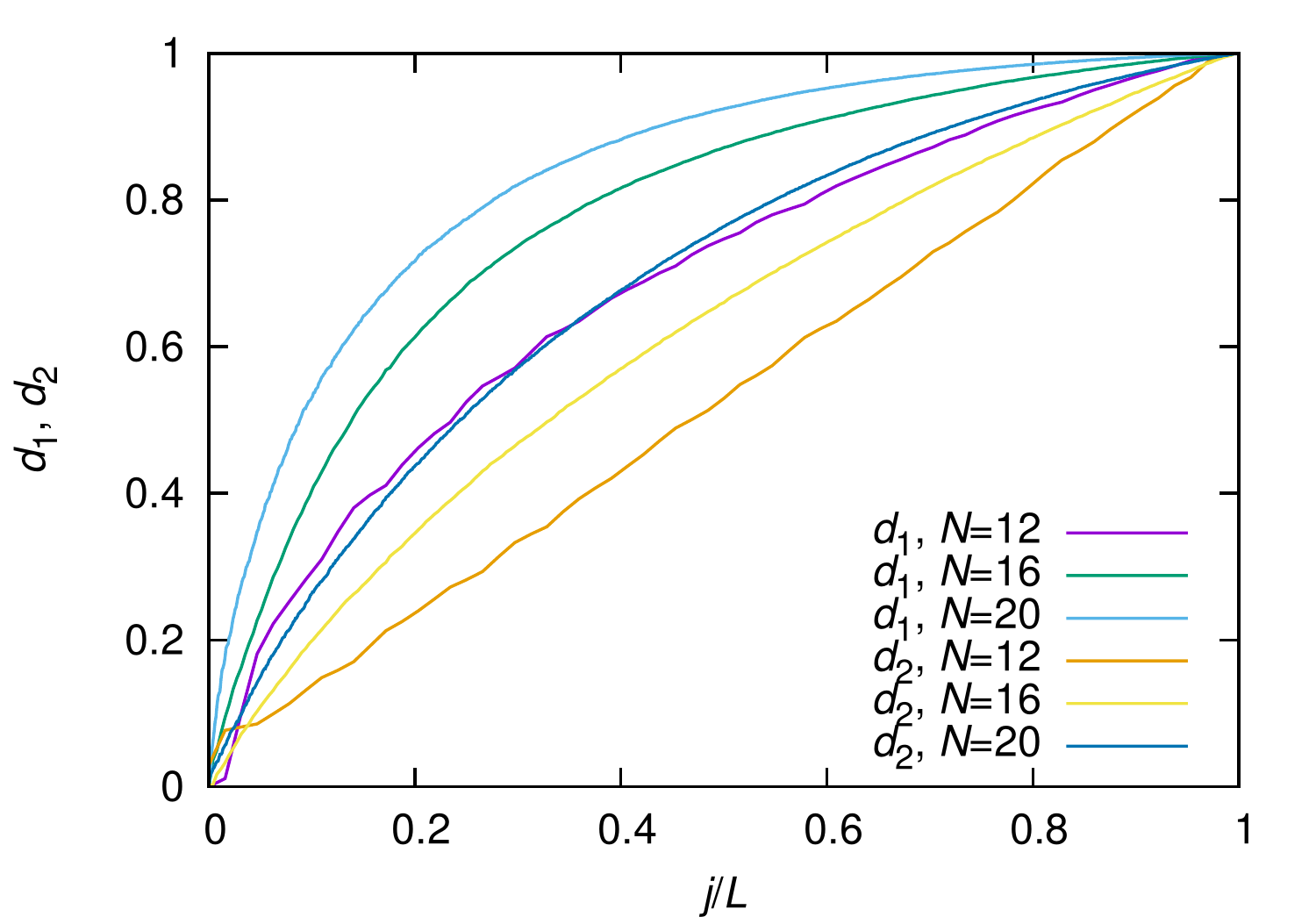}}
\rotatebox{-90}{
\includegraphics[width=5.cm]{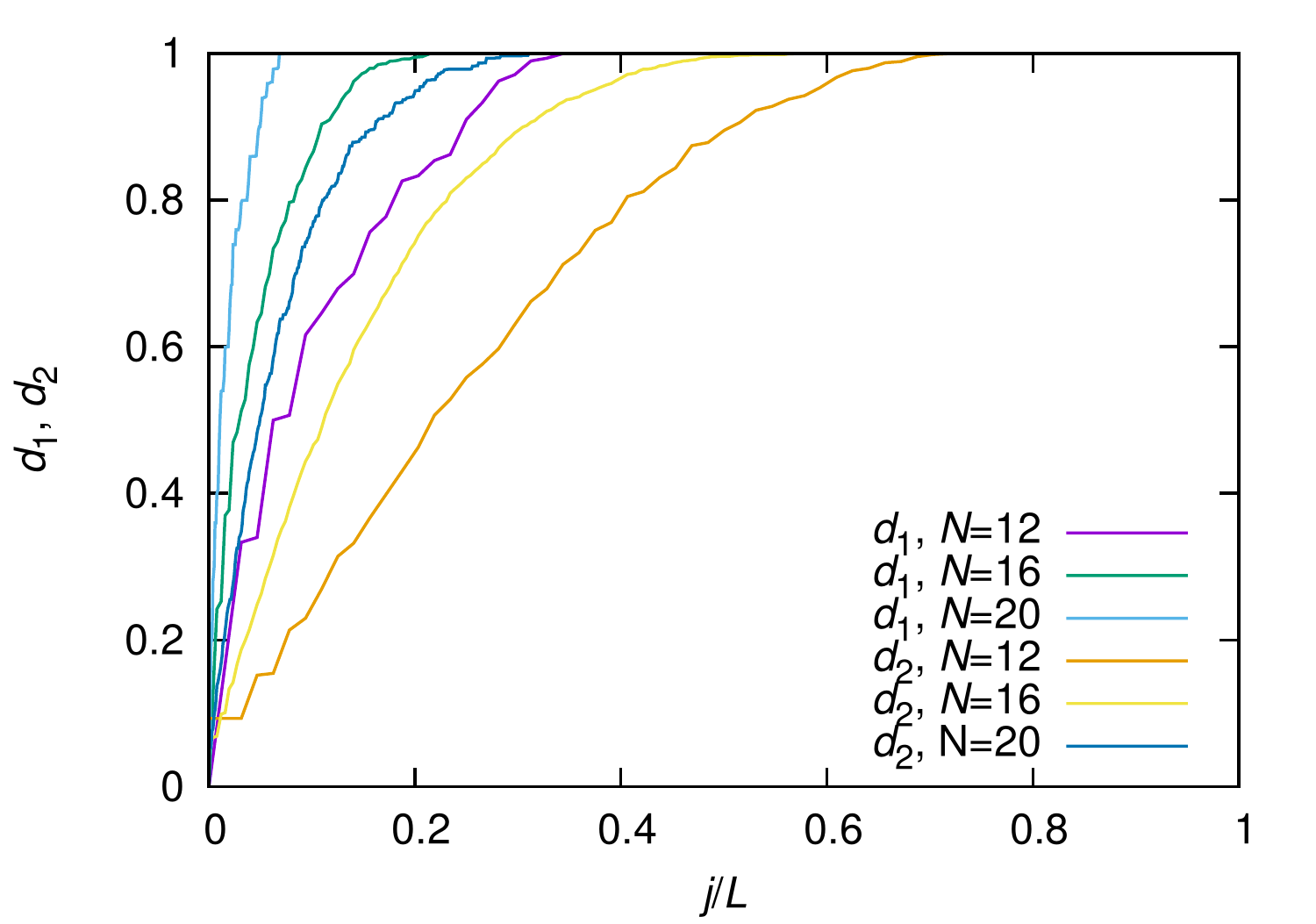}}
  \caption{
The generalized Majorana SYK model.
  $d_1$ and $d_2$ vs $j/L$. Averages over 50 samples ($N=12,16$) and 5 samples ($N=20$). [Left] $K=0.01$, [Right] $K=10$.
  }\label{overlap}
  \end{center}
\end{figure}

\hspace{0.51cm}
When an energy eigenstate $|E\rangle$ is `perturbed' by the multiplications of $\hat{\psi}$'s,
the `perturbation' is small when the energy is still well localized around $E$ after the `perturbation'.
To see it quantitatively, we take $|E\rangle$ to be the ground state $|E_0\rangle$, and plot
$d_1(j)\equiv \frac{2}{N}\sum_{k=1}^N\sum_{i=1}^j |\langle E_i|\hat{\psi}_k|E_0\rangle|^2$
and $d_2(j)\equiv \frac{4}{N}\sum_{k=1}^N\sum_{i=1}^j |\langle E_i|\hat{\psi}_k\hat{\psi}_{k+1}|E_0\rangle|^2$ for $N=12, 16$ and $20$  in Fig.~\ref{overlap}.
Here the energy eigenvalues are ordered as $E_0\le E_1\le\cdots\le E_{L-1}$ ($L=2^{N/2}$).
We can see rather large deviations of $d_1(j)$ and $d_2(j)$ from 1 even at large values of $j$,
which means that $\hat{\psi}_k|E_0\rangle$ and $\hat{\psi}_k\hat{\psi}_{k+1}|E_0\rangle$ involve large contributions
from the excited states. When $K$ is close to zero, they are almost uniform superposition of all eigenstates.
Therefore, with the resources we used for this paper, it is difficult to study physics at different energy scales separately,
especially when $K$ is close to zero.

According to Ref.~\cite{Garcia-Garcia:2017bkg,Nosaka:2018iat},
the model under consideration is integrable at sufficiently low temperature, when $K>0$ and $N=\infty$.
As we have seen already, it is hard to study the properties of the integrable phase and chaotic phase separately.
However by varying the value of $K$ we can change the numbers of integrable and chaotic states;
as $K$ becomes larger, more energy eigenstates belong to the integrable sector.
Hence we can learn about the difference between two phases by looking at the way the property of the mixture changes.
\subsubsection*{XXZ}\label{sec:perturbation_size_XXZ}
\hspace{0.51cm}

\begin{figure}[htbp]
\begin{center}
\includegraphics[trim=100 100 250 250,width=10cm]{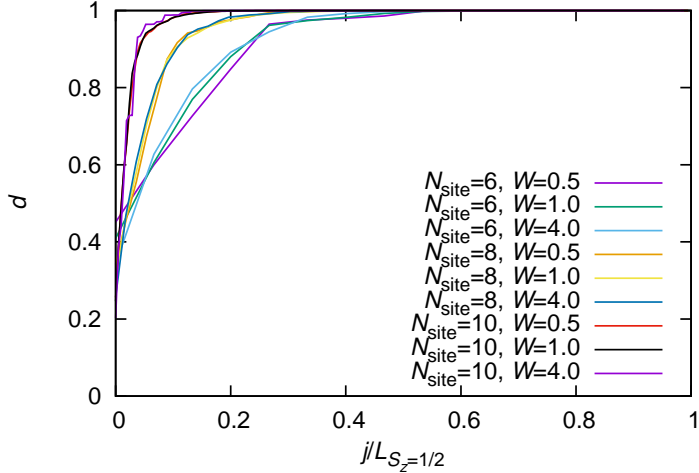}
\end{center}
\caption{
The XXZ model.
$d\equiv
\frac{1}{N_{\rm site}}
\sum_{k=1}^{N_{\rm site}}
\sum_{i=0}^j
|\langle E_{i,1/2}|
\sigma_k^+
|E_{0,0}\rangle|^2$.
Averages over 100 samples ($N_{\rm site}=6, 8$) and 5 samples ($N_{\rm site}=10$).
The horizontal axis is $j/L_{S_z=1/2}$, where $L_{S_z=1/2}$ is the dimension of 
$S_z= 1/2$ Hilbert space.
}
\label{fig:overlap_XXZ}
\end{figure}

In order to estimate the size of the perturbation, we calculated
\begin{eqnarray}
d\equiv
\frac{1}{N_{\rm site}}
\sum_{k=1}^{N_{\rm site}}
\sum_{i=0}^j
|\langle E_{i,1/2}|
\sigma_k^+
|E_{0,0}\rangle|^2.
\end{eqnarray}
Here $|E_{i,s}\rangle$ is the energy eigenstate in the total spin $s$ sector, ordered as $E_{0,s}\le E_{1,s}\le\cdots$.
The results are shown in Fig.~\ref{fig:overlap_XXZ}. We can see that the perturbations are actually small,
unlike the case of SYK.

\subsection{Relations to Other Approaches}
\hspace{0.51cm}
Usually the Lyapunov exponent is defined in terms of OTOC,
\begin{eqnarray}
\langle
[\hat{V}(t), \hat{W}(0)]^2
\rangle_{\beta}
\sim
e^{2\lambda^{\rm (OTOC)}t},
\end{eqnarray}
where $\hat{V}$ and $\hat{W}$ are arbitrary local operators and $\langle\ \cdot\ \rangle_{\beta}$
stands for the thermal average with temperature $T=\beta^{-1}$.
The idea is that the Lyapunov growth with the largest exponent can be captured by
$\hat{M}_{ij}$ with any combination of $\hat{z}_i$ and $\hat{\Pi}_{j}$, and we do not need to
take a specific basis.
By assuming the eigenstate thermalization hypothesis, the thermal average should be indistinguishable
from the expectation value taken in a typical energy eigenstate, and then
the components of $L_{ij}^{(\phi)}$ (with $\vert\phi\rangle$ taken to be a typical energy eigenstate)
should grow as $e^{2\lambda_i t}$, and the largest eigenvalue should grow as $e^{2\lambda_Nt}$.
At late time, the largest exponent dominates the growth.
Hence the largest Lyapunov exponent obtained by using our definition should be the same as the Lyapunov exponent
defined from OTOC at sufficiently late time.

One might consider $\sum_i L_{ii}^{(\phi)}(t)$ instead of $\langle [\hat{V}(t), \hat{W}(0)]^2\rangle_{\beta}$; if a given theory were gauge theory,
it would be a gauge invariant quantity. Actually the spectrum $\lambda_i$ can naturally capture the sub-leading contributions in the Lyapunov growth,
because
\begin{eqnarray}
\sum_i L_{ii}^{(\phi)}(t)
=
\sum_i s_i^2
=
\sum_i e^{2\lambda_i t}.
\label{eq:usual_Lyap_growth_OTOC}
\end{eqnarray}
In Ref.~\cite{Roberts2018}, an operator which is essentially the same as $\sum_i L_{ii}^{(\phi)}(t)$ has been considered,
and the interpretation as the growth of a size of local operator has been explained.

Note that, even when one is interested only in the largest Lyapunov exponent,
a use of the spectrum can have technical gain for numerical calculations.
If one uses the usual OTOC to define the Lyapunov exponent as $e^{2\lambda^{\rm (OTOC)}t}\equiv\frac{1}{N}\sum_i L_{ii}^{(\phi)}(t)=\frac{1}{N}\sum_i e^{2\lambda_i t}$,  the contribution from the smaller exponents can have non-negligible contribution at early time.
Numerically, it is not easy to study sufficiently late time where the largest exponent dominates.
By calculating the spectrum, it is possible to extract the largest exponent even at early time.
We will demonstrate this in Sec.~\ref{sec:Lyapunov_growth} (Fig.~\ref{Fig:lambda_OTOC-ve-lambda_N} and Fig.~\ref{fig:XXZ-lyap-OTOC}).

\subsubsection*{Lyapunovian}
\hspace{0.51cm}
A related notion called the `Lyapunovian' was recently proposed in Ref.~\cite{rozenbaum2018universal}. Suppose we have a system with one $\hat{x}$ and one $\hat{\Pi}$.
By using energy eigenstates $|E_m\rangle$, we can make the `Lyapunovian matrix'
\begin{eqnarray}
\langle E_m|
[\hat{x}(t),\hat{\Pi}(0)]^2
|E_n\rangle.
\end{eqnarray}
Then one can study the eigenvalue statistics of this matrix.
A natural counterpart for the many-body case is
\begin{eqnarray}
\langle E_m|
[\hat{z}_i(t),\hat{\Pi}_j(0)]^2
|E_n\rangle.
\end{eqnarray}
This is complementary to our approach, in which we tried to see the fixed-energy physics for each reference state. Note also that the size of this Lyapunovian matrix in the many-body case is exponentially large, unlike our approach where the size of the matrix is of order the number of degrees of freedom.

We can also consider a hybrid,
\begin{eqnarray}
L_{im;jn}(t)
\equiv
\langle E_m|\hat{M}^\ast_{ki}(t)\hat{M}_{kj}(t)|E_n\rangle.
\label{def-Lij-hybrid}
\end{eqnarray}
Then there is no ambiguity associated with a choice of a reference state $|\phi\rangle$.
It would be interesting to study the properties of the spectrum obtained from this matrix.

\subsubsection*{Lyapunov spectrum from projection}
\hspace{0.51cm}
Another recently proposed approach due to Ref.~\cite{hallam2018lyapunov} considers projecting the many-body Schrodinger equation onto a subspace of states in the full Hilbert space, specifically a set of low-entanglement matrix product states. The resulting projected dynamics can be viewed as a classical nonlinear dynamical system with a symplectic structure. As such, it can exhibit classical chaos and has a notion of Lyapunov spectrum. This auxiliary classical problem gives another way of associating a Lyapunov spectrum with an arbitrary quantum system.

Based on the results of Ref.~\cite{Xu2018}, we expect that this projected low-entanglement dynamics can accurately capture the long-distance early growth of OTOCs. Hence, one might expect that part of their spectrum agrees with our definition, although we have not checked this. However, there has also been some work advocating caution with such an approach~\cite{Kloss2017}. More generally, it is not clear to us how their full spectrum relates to the spectrum we defined. It would be interesting to determine if their spectrum also exhibits random matrix statistics, as suggested by our results.

\section{Lyapunov Growth}\label{sec:Lyapunov_growth}
\hspace{0.51cm}
In this section, we present numerical results for the Lyapunov growth in SYK and XXZ models. We then compare the Kolmogorov-Sinain entropy growth to the entanglement entropy growth inspired by classical analogy of KS entropy and coarse grained entropy.

\subsection{Lyapunov Growth in SYK}
\hspace{0.51cm}

\subsubsection{The largest exponent vs $\lambda^{\rm (OTOC)}$}\label{sec:SYK-largest-exponent}
\begin{figure}[htbp]
\begin{center}
\includegraphics[width=8cm]{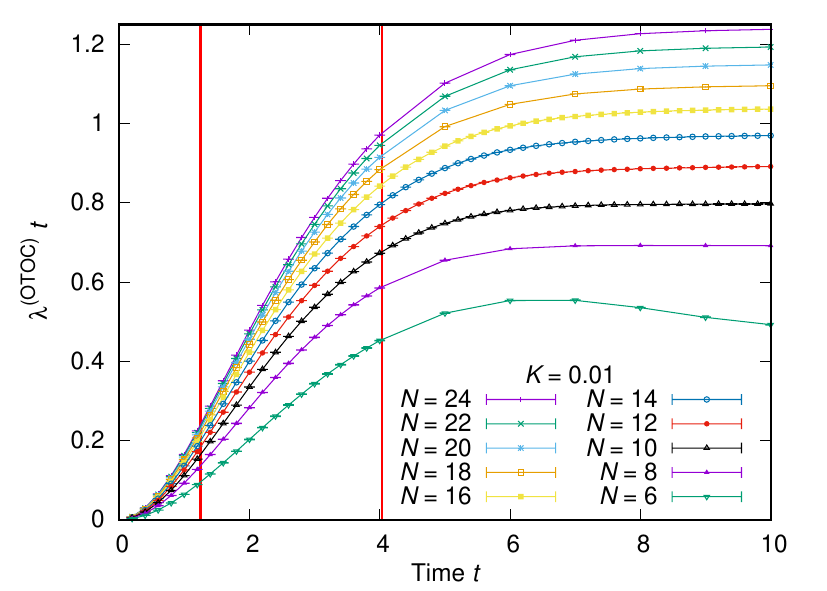}
\includegraphics[width=8cm]{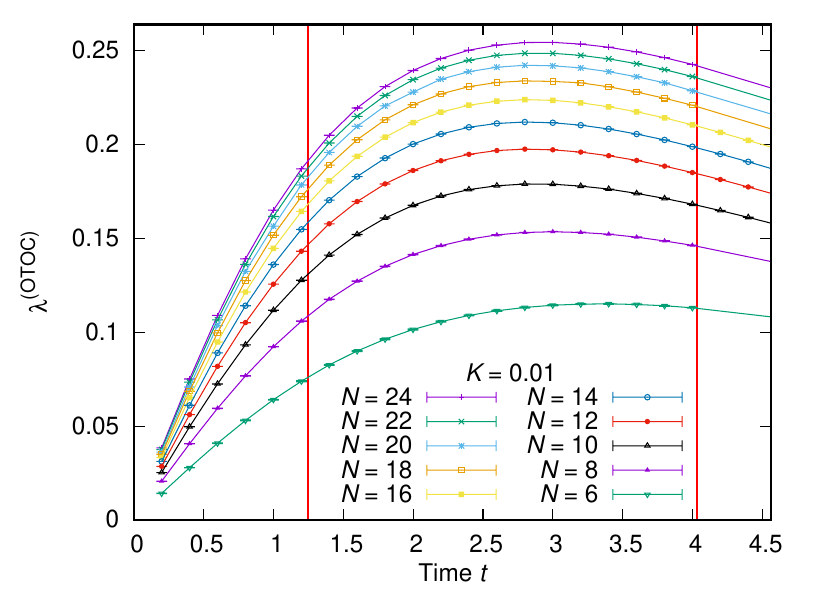}
\includegraphics[width=8cm]{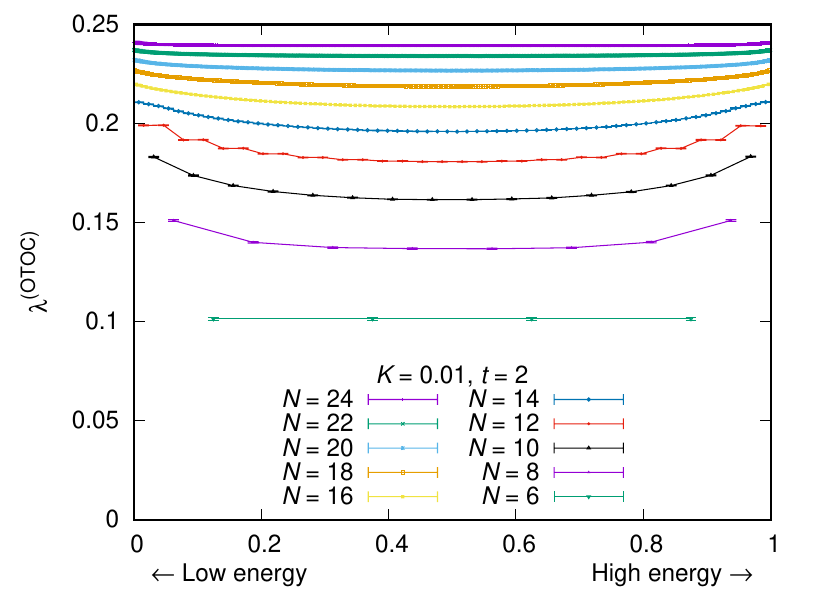}
\end{center}
\caption{
The generalized SYK model with $K=0.01$.
[Left] The Lyapunov growth from OTOC, the almost linear dependence on time $t$ of $\lambda^{(\mathrm{OTOC})}t=\frac12\log\left(\frac{1}{N}\sum_{i=1}^N e^{2\lambda_i t}\right)$.
10 samples for $N=24, 22$ and 1000 samples for $N=20, 18, \ldots, 6$ are used.
Vertical lines correspond to 20\% and 80\% for $N=20$.
[Middle] The Lyapunov exponent estimated from OTOC, $\lambda^{\mathrm{(OTOC)}}\equiv\frac{1}{2t}\log\left(\frac{1}{N}\sum_{i=1}^N e^{2\lambda_i t}\right)$.
[Right] At each $N$, the exponent $\lambda^{\mathrm{(OTOC)}}$ at $t=2$ is shown as the function of the energy $E_i$.
  The horizontal axis is $(i+1/2)/L$, so that the left and right corresponds to low and high energies.}\label{fig:OTOC}
\end{figure}

\begin{figure}[htbp]
\begin{center}
\includegraphics{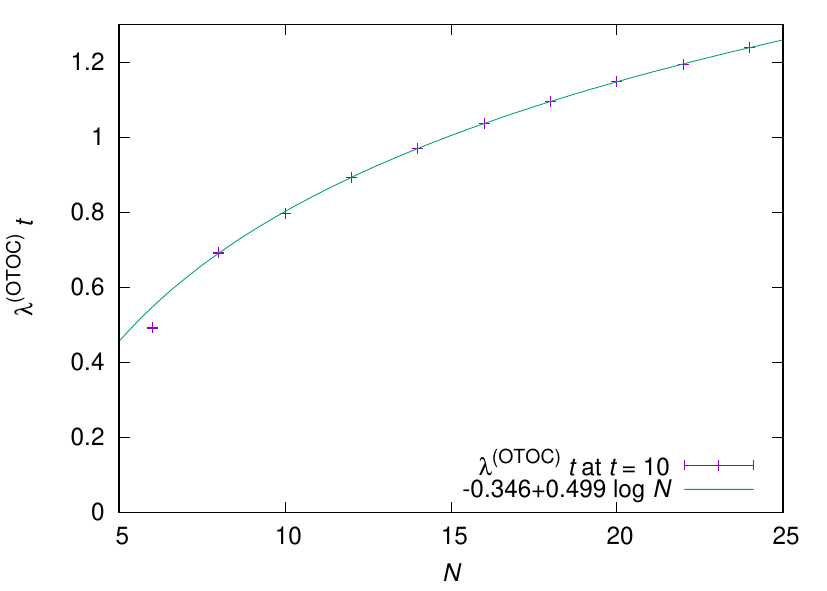}
\end{center}
\caption{Plot of the value of $\lambda^\mathrm{(OTOC)} t$ at $t=10$ along with a fit with a linear function of $\log N$ for the generalized SYK model with $K=0.01$.}
\label{OTOC-lyap-Nm6-24-t10-K001-logfit.pdf}
\end{figure}

\hspace{0.51cm}
Let us start with the `usual' OTOC \eqref{eq:usual_Lyap_growth_OTOC},
which is
\begin{eqnarray}
e^{2\lambda^{\rm (OTOC)}t}
\equiv
\frac{1}{N}
\sum_{i=1}^N L_{ii}^{(\phi)}(t)
=
\frac{1}{N}
\sum_{i,j=1}^N
\left\langle\phi|
\{\psi_i(t),\psi_j(0)\}^2
|\phi\right\rangle
=
\frac{1}{N}
\sum_{i=1}^N e^{2\lambda_i t}.
\label{eq:lambda_OTOC}
\end{eqnarray}
In the left panel of Fig.~\ref{fig:OTOC}, we plot $\lambda^{\rm (OTOC)}t$ at $\beta=1/T=0$.
We can see the exponential growth followed by the saturation at $\lambda^{\rm (OTOC)}t\sim \log N$
as in Fig.~\ref{OTOC-lyap-Nm6-24-t10-K001-logfit.pdf}.\footnote{
The fit value is close to $e^{2\lambda^{\rm (OTOC)}t}=\frac{N}{2}$.
This is value can be explained as follows.
At late time, all $\left\langle\phi|\{\psi_i(t),\psi_j(0)\}^2|\phi\right\rangle$ in \eqref{eq:lambda_OTOC} give the same contribution.
Each of them contains four terms, $\left\langle\phi|\psi_j(0)\psi_i(t)\psi_i(t)\psi_j(0)|\phi\right\rangle$,
$\left\langle\phi|\psi_i(t)\psi_j(0)\psi_j(0)\psi_i(t)|\phi\right\rangle$,
$\left\langle\phi|\psi_i(t)\psi_j(0)\psi_i(t)\psi_j(0)|\phi\right\rangle$
and
$\left\langle\phi|\psi_j(0)\psi_i(t)\psi_j(0)\psi_i(t)|\phi\right\rangle$.
The first two terms are $\frac{1}{4}$, while the latter two terms are suppressed at large $N$.
Hence $e^{2\lambda^{\rm (OTOC)}t}\to \frac{1}{N}\sum_{i,j=1}^N 2\times \frac{1}{4} = \frac{N}{2}$ up to a small correction.
 }
Two red vertical lines show the times at which this growth reached 20\% and 80\% for $N=20$.
Between them, the slope is approximately constant for each $N$. We can see it more clearly in the middle panel,
where $\lambda^{\rm (OTOC)}$ is shown.
At $6\le N\le 24$, the exponent $\lambda^{\rm (OTOC)}$ changes substantially with $N$,
and it is not easy to take the large-$N$ limit.
In the right panel, we have shown the energy dependence of $\lambda^{\rm (OTOC)}$,
by taking $|\phi\rangle$ to be the energy eigenstates.
Strangely, the exponent in the ground state is larger than the one in the excited states.
As we will explain later, this is because the finite-$N$ effect is large.

\begin{figure}[htbp]
\scalebox{1.}{\includegraphics{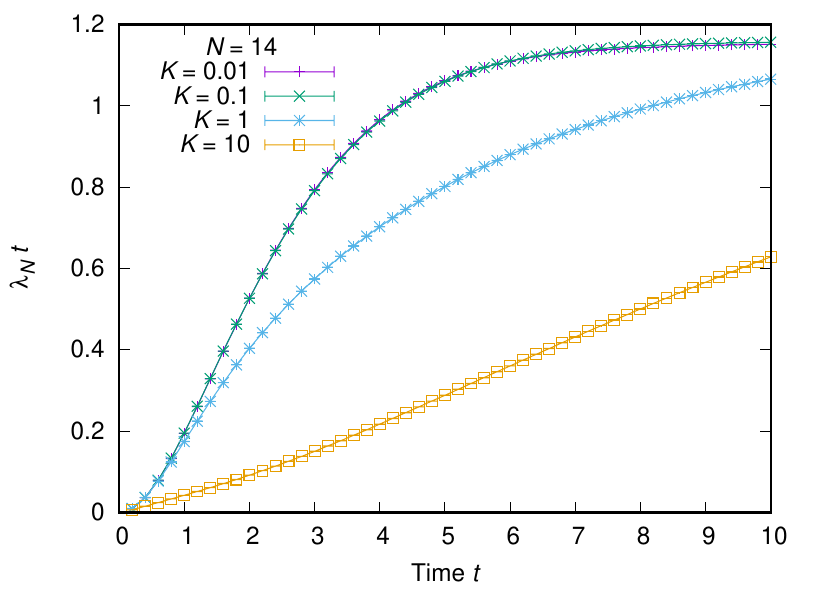}}
\scalebox{1.}{\includegraphics{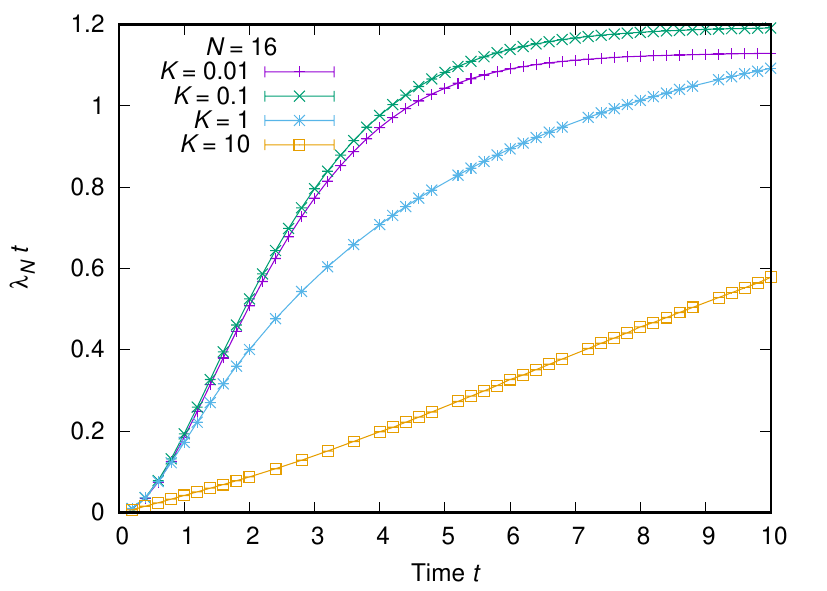}}\\
\scalebox{1.}{\includegraphics{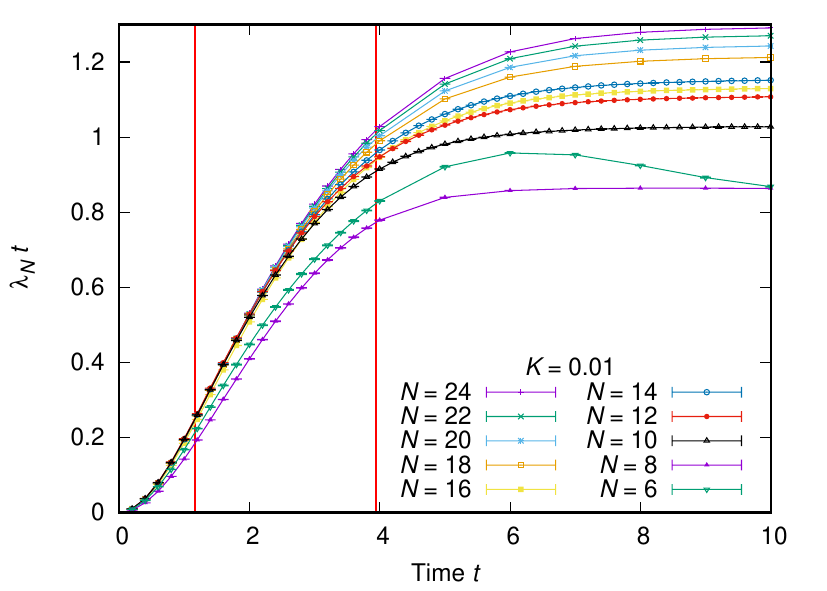}}
\scalebox{1.}{\includegraphics{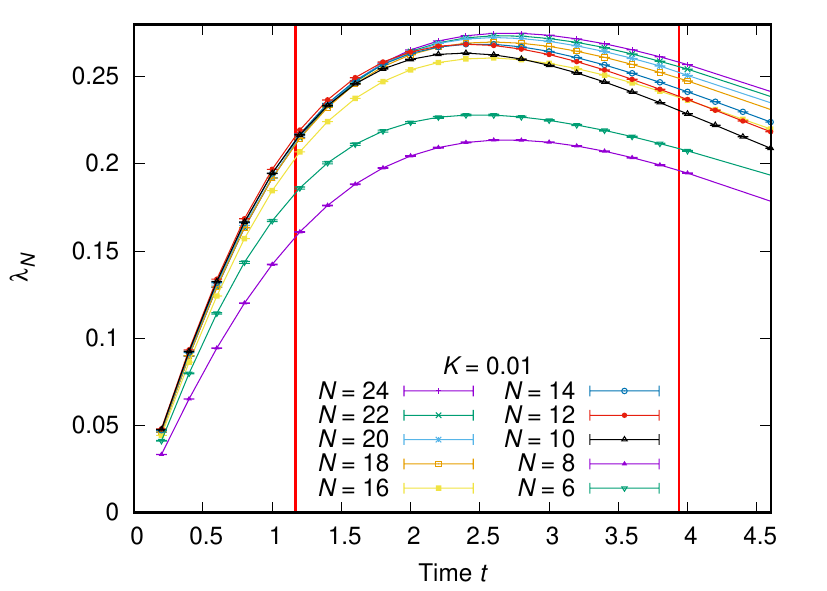}}
\caption{
The average of the largest Lyapunov exponents obtained using all eigenvectors of the generalized SYK model.
[Upper-left] $\lambda_N t$ vs $t$ for $N=14$, various $K$.
[Upper-right] $\lambda_N t$ vs $t$ for $N=16$, various $K$.
[Lower-left] The $N$ dependence of $\lambda_N t$ for $K=0.01$. The vertical lines corresponding to 20\% and 80\% of the plateau value (at $t=10$) of $\lambda_\mathrm{L} t$ for $N=20$ are also shown.
[Lower-Right] The $N$ dependence of $\lambda_N$ for $K=0.01$.
}\label{fig:SYK-Lyapunov-growth-1}
\end{figure}

\begin{figure}[htbp]
\begin{center}
\scalebox{1.}{\includegraphics{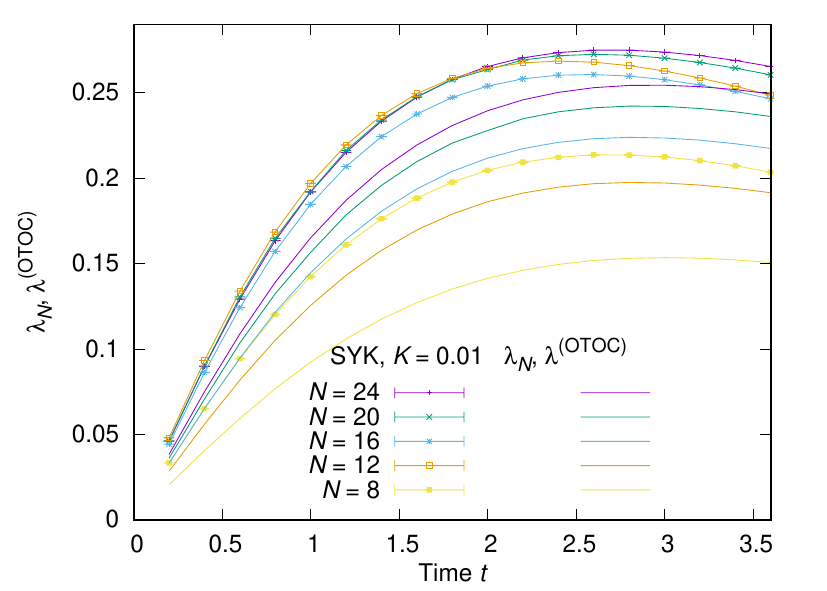}}
\end{center}
  \caption{
$\lambda_N$ (with data points and error bars) and $\lambda^{\rm (OTOC)}$ (with the same color as $\lambda_N$ but data points omitted)
plotted against time $t$
for the generalized Majorana SYK model with $K=0.01$.
  }\label{Fig:lambda_OTOC-ve-lambda_N}
\end{figure}

Next let us see the Lyapunov spectrum obtained from our definition.
In the first two panels of Fig.~\ref{fig:SYK-Lyapunov-growth-1}, we have plotted $\lambda_N t$
for several values of $K$ and $N=14, 16$ at $\beta=0$; we can see the exponential growth followed by the saturation.
The third and fourth panels in Fig.~\ref{fig:SYK-Lyapunov-growth-1} show the $N$ dependence of $\lambda_N t$ and $\lambda_N$
for $K=0.01$. Compared to $\lambda^{\rm (OTOC)}$, the value at each $N$ is larger (see Fig.~\ref{Fig:lambda_OTOC-ve-lambda_N}), because $\lambda^{\rm (OTOC)}$
contains the `contamination' from the smaller exponents $\lambda_1, \cdots,\lambda_{N-1}$. As we will see shortly,
the finite-$N$ corrections to the smaller exponents are larger than the one to the largest exponent.
For this reason, $\lambda^{\rm (OTOC)}$ depends more severely on $N$.
Note also that the $N$-dependence of $\lambda_N$ is not smooth, but rather it shows sensitive dependence on $N$ mod 8,
which suggests that $\lambda_N$ captures the finer detail of the theory at finite $N$.

\begin{figure}[htbp]
\begin{center}
\scalebox{0.8}{
\includegraphics{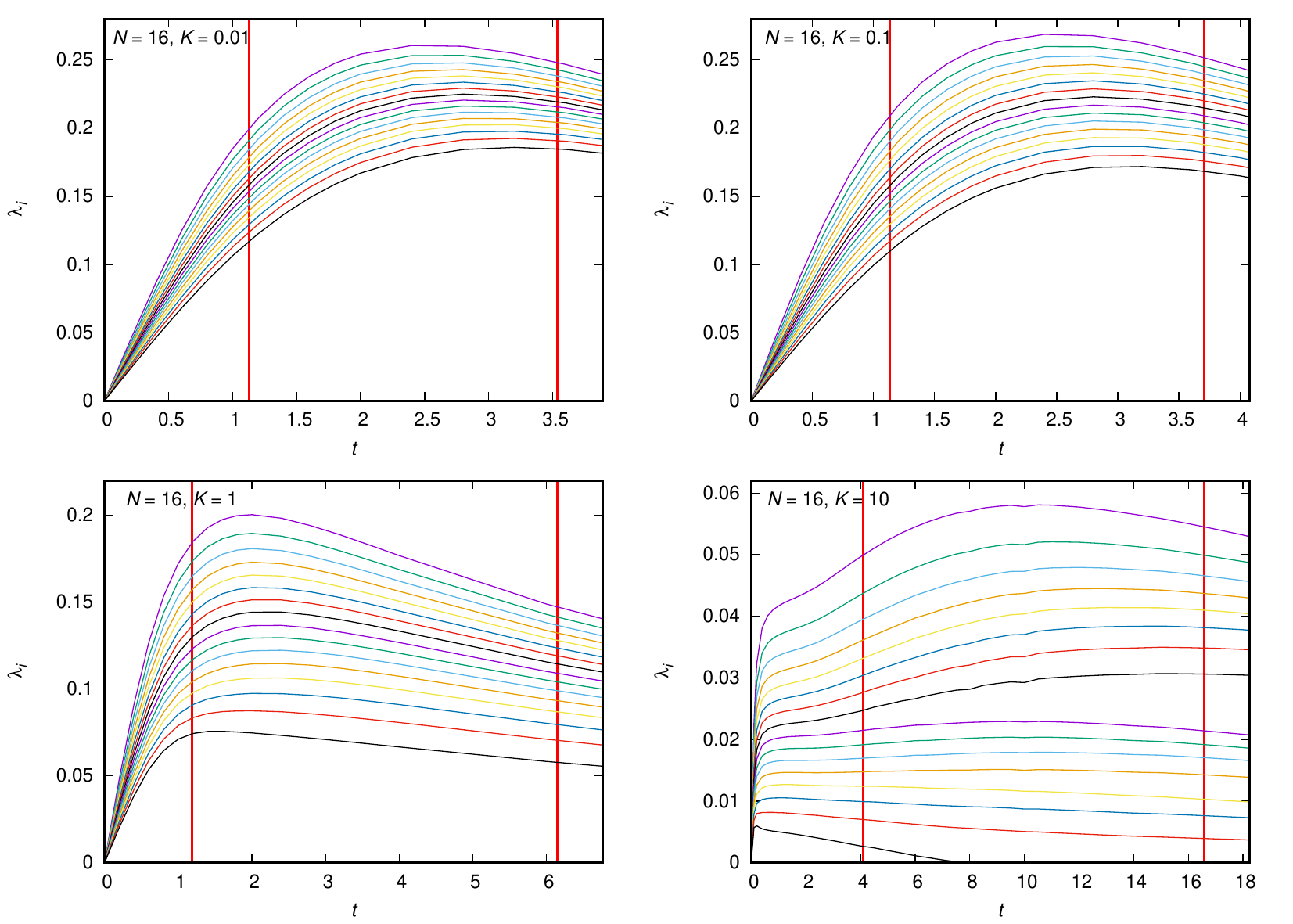}}
\end{center}
  \caption{
The Lyapunov spectrum of the generalized Majorana SYK model, $N=16$, $\beta=0$, with $K=0.01, 0.1, 1, 10$.
For $K\gtrsim 10$, the separation between the larger eight exponents and the lower eight exponents become increasingly clear for larger $t$.
The times at which the Lyapunov growth ($\lambda_N t$ in our notation)
reaches $20~\%$ and $80~\%$ of its plateau value are shown by red vertical lines.
We can see near-constant behavior of $\lambda_N$ there.
  }\label{fig:SYK-Lyapunov-spectrum-beta=0}
\end{figure}

In Fig.~\ref{fig:SYK-Lyapunov-spectrum-beta=0}, all the exponents $\lambda_i$ are shown for $N=16$. When $K$ is small, all the exponents are positive. For larger $K$ ($K\gtrsim 10$), beyond $t\gtrsim1$, a gap emerges between the larger half and smaller half of the exponents,
and the density distribution of the lower eight exponents become increasingly sharper and get closer to zero.
Two red vertical lines represent the 20\% and 80\% saturation of $\lambda_N t$.
Between them the exponents are almost constant.

In Fig.~\ref{Fig:lambda_OTOC-ve-lambda_N}, the largest exponent $\lambda_N$
is compared with $\lambda^\mathrm{(OTOC)}$.
As explained around Eq.~\eqref{eq:usual_Lyap_growth_OTOC},
$\lambda^\mathrm{(OTOC)}$ is contaminated by the smaller exponents;
indeed, we can see clear difference in the plot.
It is interesting to note that the difference between $\lambda_N$ and $\lambda^\mathrm{(OTOC)}$ becomes smaller as $N$ increases.
In Sec.~\ref{sec:fastest-entropy-generator-conjecture}, we will study this point further.

\begin{figure}[htbp]
\begin{center}
\includegraphics[width=5.cm]{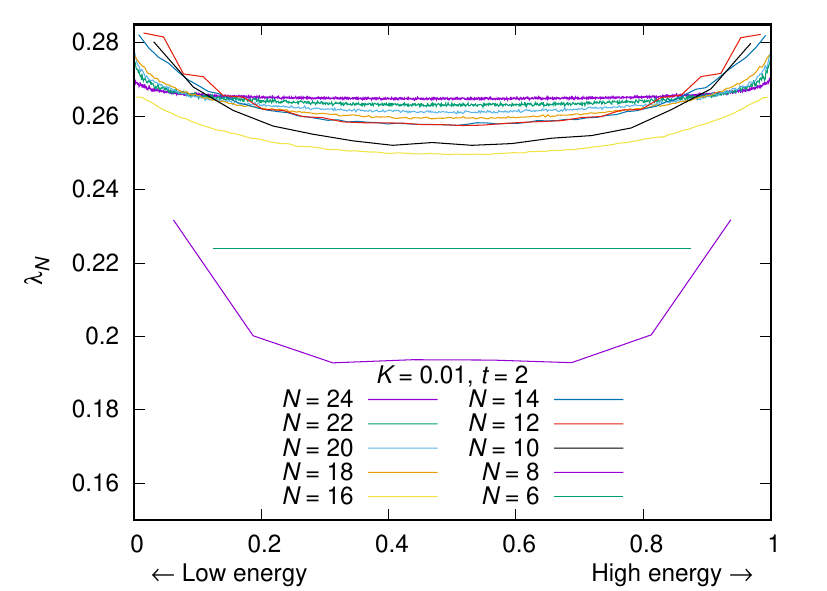}
\includegraphics[width=5.cm]{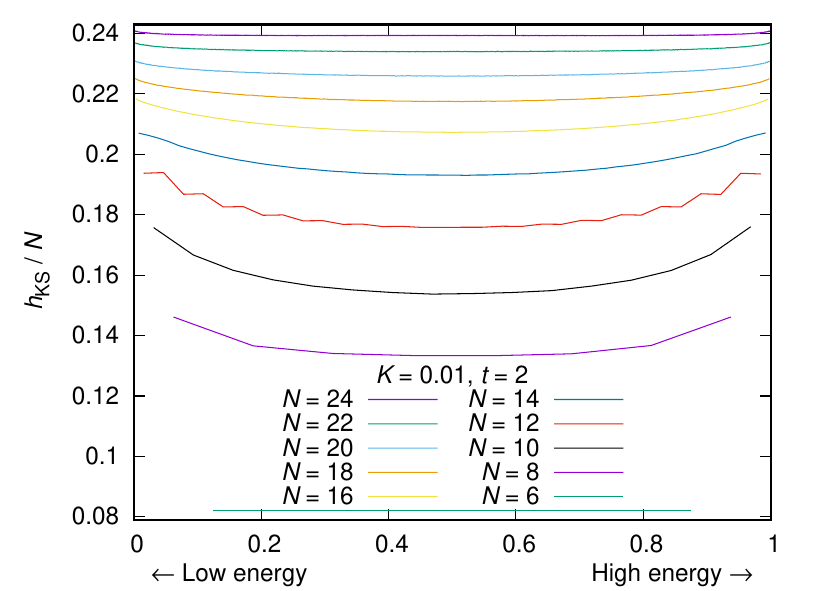}
\includegraphics[width=5.cm]{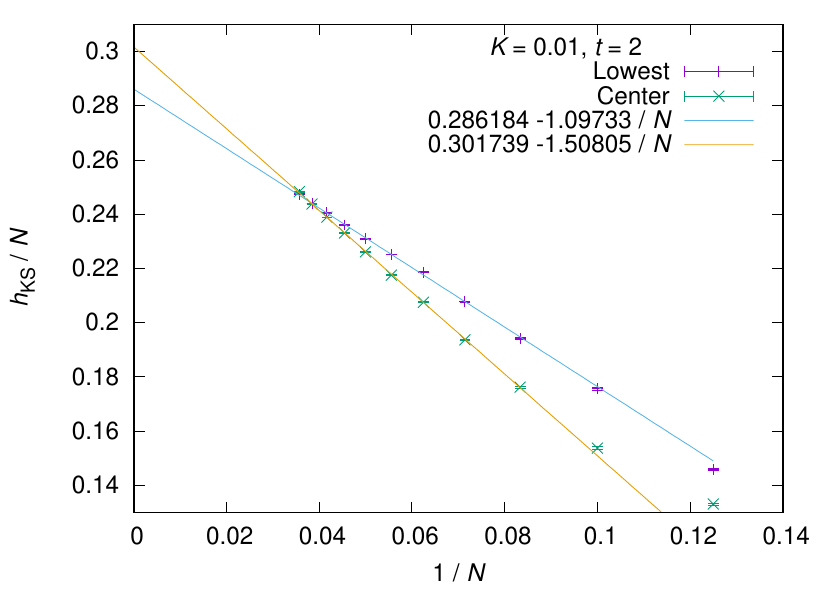}
\\
\end{center}
  \caption{
The generalized Majorana SYK model with $K=0.01$.
  [Left] At each $N$, the largest Lyapunov exponent $\lambda_N$ at $t=2$ is shown as the function of the energy $E_i$.
  The horizontal axis is $(i+1/2)/L$, so that the left and right corresponds to low and high energies.
[Middle] $h_\mathrm{KS}/N$ vs energy, at $t = 2$.
[Right] $h_\mathrm{KS}/N$  against $1/N$, for the ground state and the center of the energy spectrum.}
\label{KS-entropy}
\label{fig:SYK-Lyapunov-N-dependence}
\end{figure}

Fig.~\ref{fig:SYK-Lyapunov-N-dependence} is made in order to see the effect of the choice of the reference state.
The left panel is the energy vs the largest exponent $\lambda_N$ at $N\le 24$.
There are two peculiar points here (note that we observed the same for $\lambda^\mathrm{(OTOC)}$ in Fig.~\ref{fig:OTOC}):
The growth rate does not seem to depend heavily on the choice of the energy eigenstate.
Even when we take $|\phi\rangle$ to be the ground state, we can still see the exponential growth.
Furthermore, the ground state gives faster growth.
Seemingly it has a tension with the large-$N$ result at low temperature, $\lambda=2\pi T$.
Presumably this is because the values of $N$ studied here are so small that the energy excitation caused by operator $\hat{M}$ is not negligible,
as we have seen in Sec.~\ref{sec:perturbation_size}.
At least, as we can see from the right panel, the exponents calculated by using the ground state $|E_0\rangle$
become smaller than the ones obtained from the state at the center of the energy spectrum $|E_{L/2}\rangle$ at $N\gtrsim 26$.
Still, our numerics is not good enough to show whether $\lambda_N$ defined from the ground state vanishes in the large-$N$ limit,
as expected from both the usual intuitive picture and the MSS bound.

\subsubsection{Kolmogorov-Sinai and Entanglement Entropy}\label{sec:SYK-EE-vs-KS}
\hspace{0.51cm}
For classical systems, the Kolmogorov-Sinai entropy $h_\mathrm{KS}$ is defined as the sum of all positive Lyapunov exponents.
We use an analogous definition here for the quantum Lyapunov exponents we have defined in this paper; see Eq.~\eqref{KS-entropy-def}.

\subsubsection*{Kolmogorov-Sinai Entropy (KS)}
\hspace{0.51cm}
In the middle panel of Fig.~\ref{KS-entropy}, the energy dependence of $h_{KS}/N$ is shown.
At small $N$, the curve is concave due to the finite-$N$ effect.
As $N$ becomes large, it will become convex as we can see from the right panel of Fig.~\ref{KS-entropy}.

\subsubsection*{Entanglement Entropy (EE)}
\hspace{0.51cm}
Let us introduce Dirac fermions $\hat{c}_k=\frac{\hat{\psi}_{2k-1}+\sqrt{-1}\hat{\psi}_{2k}}{\sqrt{2}}$
($k=1, 2, \ldots, N/2$).
Then we can label the states by using the excitation number 0 or 1 for $\hat{c}_k$.
We use the state $|00\cdots 0\rangle$ to calculate the entanglement entropy.
(Note this is not the ground state of the Hamiltonian.)
We factorize the Hilbert space to be ${\cal H}_{|A|}\times {\cal H}_{N/2-|A|}$,
where ${\cal H}_{|A|}$ and ${\cal H}_{N/2-|A|}$ are generated by acting $\hat{c}^\dag_1,\cdots,\hat{c}^\dag_{|A|}$
and $\hat{c}^\dag_{|A|+1},\cdots,\hat{c}^\dag_{N/2}$ to the ground state, respectively. \footnote{Here we are defining fermions in terms of spin variables via the Jordan-Wigner transform. The spin Hilbert space then has a sensible tensor product structure and this decomposition is what we are referring to.}
Then we trace out ${\cal H}_{N/2-|A|}$.

\subsubsection*{KS vs EE}
\hspace{0.51cm}
As discussed in Sec.~\ref{sec:KSentropy}, in the classical limit the growth rate of the coarse grained entropy
should agree with the Kolmogorov-Sinai entropy $h_{\rm KS}$.
Therefore, we expect that, in quantum theories, $NS_{\rm EE}/|A|$,\footnote{Note that we are proposing to use the entanglement entropy as a quantum analog of classical course grained entropy. }
which corresponds to the coarse grained entropy of the system, and $h_{\rm KS}t$ should agree up to an additive constant.
As we have emphasized, for the values of $N$ studied in this paper,
$h_{\rm KS}$ captures contributions from almost all energy eigenstates.

\begin{figure}[htbp]
\begin{center}
\includegraphics[width=7cm]{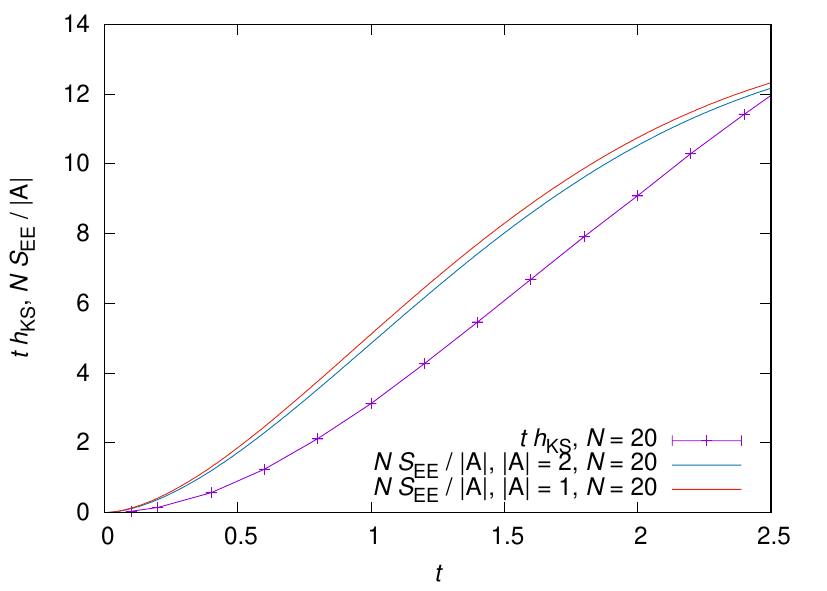}
\includegraphics[width=7cm]{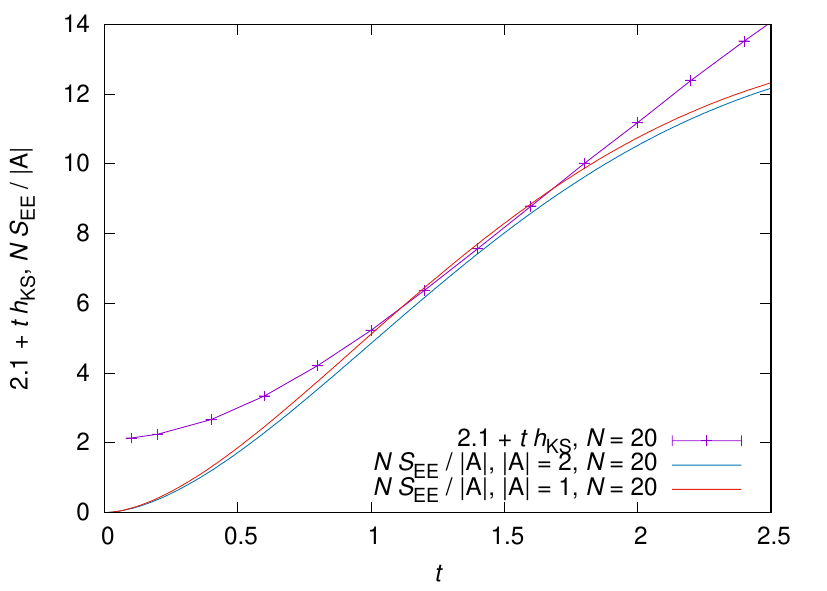}
\end{center}
\caption{
[Left] SYK model, comparison of $t h_\mathrm{KS}$ and $NS_\mathrm{EE}/|A|$.
As an initial state, the Fock vacuum annihilated by $\hat{c}_j=\frac{\hat{\psi}_{2j}-i\hat{\psi}_{2j-1}}{\sqrt{2}}$ for all $j=1,2,\cdots,\frac{N}{2}$ has been used.
[Right] Essentially the same plot, but $S_\mathrm{EE}$ is shifted by a constant.
}
\label{KS-vs-EE}
\end{figure}

In the left panel of Fig.~\ref{KS-vs-EE}, we have plotted $NS_{\rm EE}/|A|$ and $h_{\rm KS}t$
obtained from the SYK model at $\beta=0$.
At early time, they show similar growths; indeed, as we can see the right panel,
at $1\lesssim t\lesssim 2$ they agree very well just by a constant shift.\footnote{
This shift can be understood as the ambiguity of the size of the cell in Fig.~\ref{fig:coarse-graining-KS}.
}
This results are not conclusive, however, they are suggestive that the KS entropy can actually be understood as the entropy production rate in quantum systems.

\subsubsection{Fastest entropy generator?}\label{sec:fastest-entropy-generator-conjecture}
\hspace{0.51cm}
As we have seen in Sec.~\ref{sec:SYK-largest-exponent},
the difference between $\lambda_N$ and $\lambda_{\rm OTOC}$ becomes smaller as $N$ increases.
This means the largest exponent $\lambda_N$ and the smallest exponent $\lambda_1$ get close.
We can actually numerically confirm that $\lambda_N-\lambda_1$ scales as $1/N$ at $t\lesssim 2$.
This strongly suggest that the Lyapunov spectrum peaks like the delta function in the large-$N$ limit.
This is consistent with the previous analysis on the OTOC at $N=\infty$:
if the spectrum has nontrivial distribution, it can give a power law correction $t^\nu e^{\lambda_{\rm OTOC}t}$ with $\nu>0$.
But such correction has not been found \cite{Maldacena:2016hyu}.

It is interesting to compare this behavior with the usual (weakly coupled) string theory dual.
The Lyapunov spectrum of the D0-brane matrix model in the classical limit (highly stringy region)
converges to the semi-circle distribution with an $O(N^0)$ width \cite{Gur-Ari:2015rcq},
unlike the weak coupling region of SYK.
Hence the large-$N$ limit of the D0-brane theory does not by itself make the distribution peaked, unlike in the SYK model.
However the analyses on the dual gravity side including the finite coupling correction at large $N$ (the $\alpha'$ correction in gravity side) to the Lyapunov exponents \cite{Shenker2014a}
suggest that the Lyapunov exponents peak at the largest possible value (the MSS bound) at strong coupling (for example, a power-law correction has not been seen).
The same seems to be true for other quantum field theories which can be analyzed with dual gravity calculations \cite{Shenker2014a}.

Assuming this is true, both in the SYK model and quantum field theory with the usual string theory dual, all exponents saturate the MSS bound
at strong coupling and large $N$.
Therefore, both of them appear to take {\it the largest possible Kolmogorov-Sinai entropy}, or {\it the largest possible entropy production rate}.
\footnote{We repeat that we have used `Kolmogorov-Sinai entropy' to mean the sum of the Lyapunov exponents. Where this agrees with the entropy production rate
even at quantum level is a subtle issue which requires further study, although qualitative agreement has been observed as shown in Sec.~\ref{sec:SYK-EE-vs-KS}.}

For the canonical ensemble, the largest possible value of each exponent is $2\pi T$.
Apparently, when all exponents saturate this bound, the sum is maximal. Hence, it would be natural to conjecture that black hole has the largest possible KS entropy.
Note that we have the single black hole configuration (the leftmost figure in Fig.~\ref{fig:matrix-black-holes}) in mind.

For the microcanonical ensemble, it would be natural to conjecture that the entropy generation rate increases as the black hole grows,
as demonstrated for a simple case in Sec.~\ref{sec:black-hole-merger}. We can also show the same pattern for more generic initial conditions,
and we can also show that the KS entropy decreases as black hole evaporates \cite{Berkowitz:2016znt}. Therefore we conjecture that the KS entropy is maximal when all the degrees of freedom are absorbed in one black hole and thermalized.

\subsection{Lyapunov Growth in XXZ}
\hspace{0.51cm}

\begin{figure}
\centering
\includegraphics{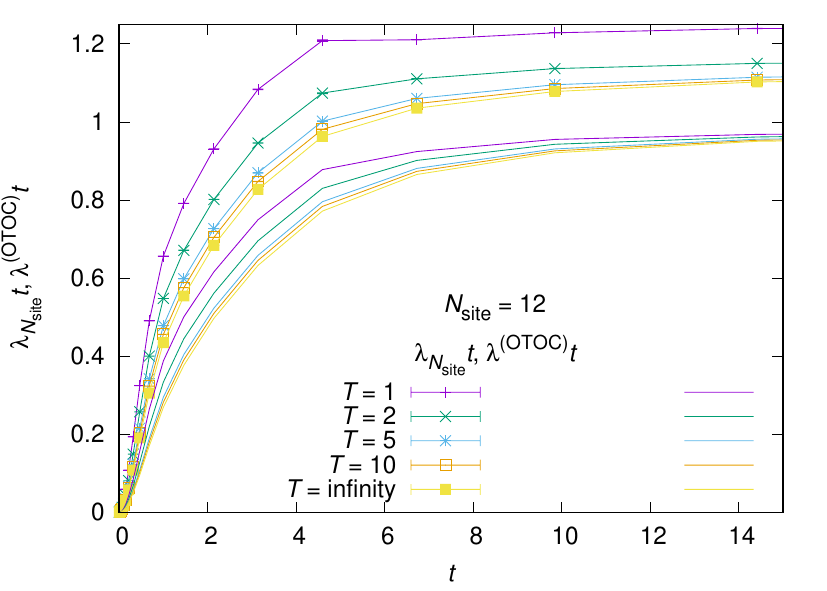}
\caption{
$\lambda_{N_\mathrm{site}} t$ (with data points and error bars) and $\lambda^{\rm (OTOC)} t$ (with the same color as $\lambda_{N_\mathrm{site}}$ but data points omitted)
plotted against time $t$
for the XXZ model with $N_\mathrm{site}=12$, $W=0.5$ and various temperatures $T$. 215 samples have been used for each data point.
}
\label{fig:XXZ-lambdaNsite-lambdaOTOC-T1-infty}
\end{figure}

Because the total $z$-spin $S^{\rm (total)}_z$ commutes with the Hamiltonian \eqref{eqn:XXZ},
we focus on the Lyapunov growth in the zero-spin sector, $\langle S^\mathrm{(total)}_z \rangle = 0$.
As we have seen in Sec.~\ref{sec:perturbation_size_XXZ}, the multiplication of $\sigma$ can actually be regarded as a small perturbation,
and hence it makes sense to study the temperature dependence, unlike the case of SYK.

In Fig.~\ref{fig:XXZ-lambdaNsite-lambdaOTOC-T1-infty} we have plotted $\lambda_{N_\mathrm{site}} t$ and $\lambda^{\rm (OTOC)} t$
as functions of $t$ for $N_\mathrm{site}=12$ and $\lambda_{N_\mathrm{site}} t$ for various temperatures.
For each $N_\mathrm{nsite}$, $\lambda^{\rm (OTOC)} t$ converges to the same value, $\frac{1}{2}\log\left(1+\frac{N_\mathrm{site}}{2}\right)$.\footnote{
For each energy eigenstate $|E\rangle$, terms of the form
$\langle E|
\sigma_{+,j}(0)\sigma_{-,i}(t)\sigma_{+,i}(t)\sigma_{-,j}(0)|E\rangle
=
||\sigma_{+,i}(t)\sigma_{-,j}(0)|E\rangle||^2
$
and
$\langle E|\sigma_{-,i}(t)\sigma_{+,j}(0)\sigma_{-,j}(0)\sigma_{+,i}(t)|E\rangle
=
||\sigma_{-,j}(0)\sigma_{+,i}(t)|E\rangle||^2$
give dominant contributions at late time.
Because we are taking $|E\rangle$ to be in the total spin zero sector, when $\sigma_{-,j}(0)$
is multiplied on $|E\rangle$, half of the terms in the $z$-spin basis --- terms with down spin at $j$-th sire --- is annihilated.
Hence $\sigma_{-,j}(0)|E\rangle$ is roughly norm $1/\sqrt{2}$, and consists of terms with $N_{\rm site}/2+1$ down spins and
$N_{\rm site}/2-1$ up spins. Then when we further multiply $\sigma_{+,i}(t)$, $(N_{\rm site}/2+1)/N_{\rm site}$ terms survive.
Hence $||\sigma_{+,i}(t)\sigma_{-,j}(0)|E\rangle||^2\simeq (N_{\rm site}/2+1)/2N_{\rm site}$.
For the same reason, $||\sigma_{-,j}(0)\sigma_{+,i}(t)|E\rangle||^2\simeq (N_{\rm site}/2+1)/N_{\rm site}$.
Hence $e^{2\lambda^{\rm (OTOC)}t}\simeq \frac{1}{N_{\rm site}}\sum_{i,j}\frac{N_{\rm site}/2+1}{2N_{\rm site}}\times 2=\frac{N_{\rm site}}{2}+1$.
}

\begin{figure}
\includegraphics{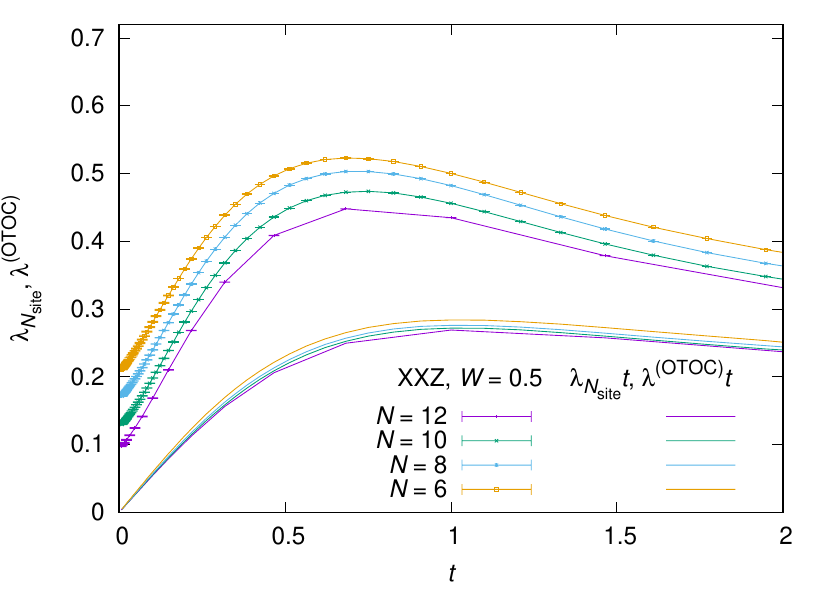}
\includegraphics{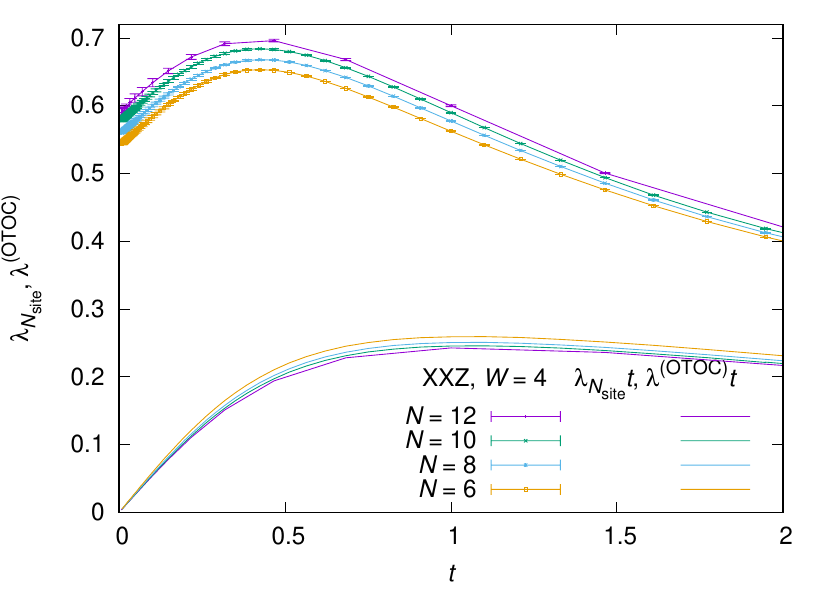}
\caption{
$\lambda_{N_\mathrm{site}}$ (with data points and error bars) and $\lambda^{\rm (OTOC)}$ (with the same color as $\lambda_{N_\mathrm{site}}$ but data points omitted) at $T=\infty$
plotted against time $t$
for the XXZ model with $N_\mathrm{site}=12$, [left] $W = 0.5$ (215 samples for $N_\mathrm{site} = 12$, more samples for $N\leq 10$) and [right] $W = 4$ (at least 102 samples for $N_\mathrm{site} = 12$, more samples for $N\leq 10$).}
\label{fig:XXZ-lyap-OTOC}
\end{figure}

The $N$-dependence of $\lambda_{N_\mathrm{site}}$ and $\lambda^\mathrm{(OTOC)}$ are shown in Fig.~\ref{fig:XXZ-lyap-OTOC} for $W=0.5$ and $W=4$.

\begin{figure}[htbp]
\includegraphics[width=5cm]{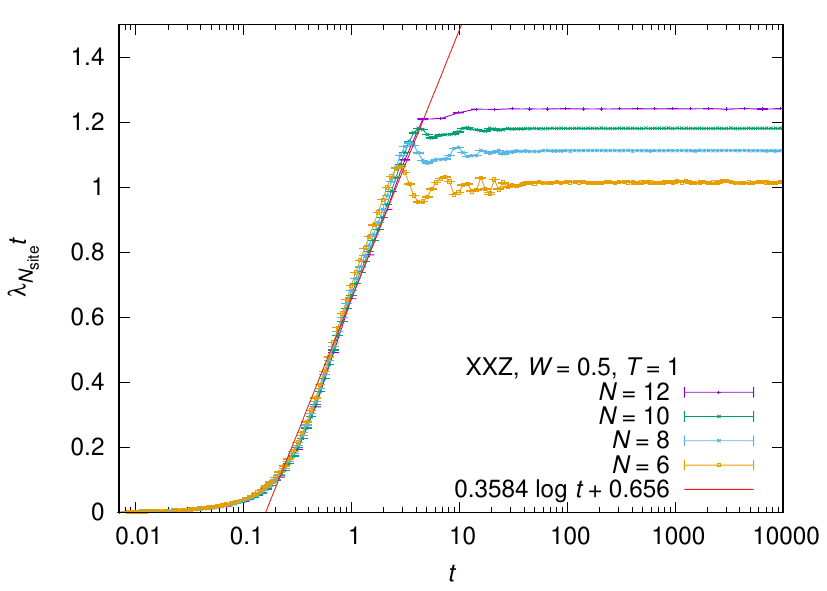}
\includegraphics[width=5cm]{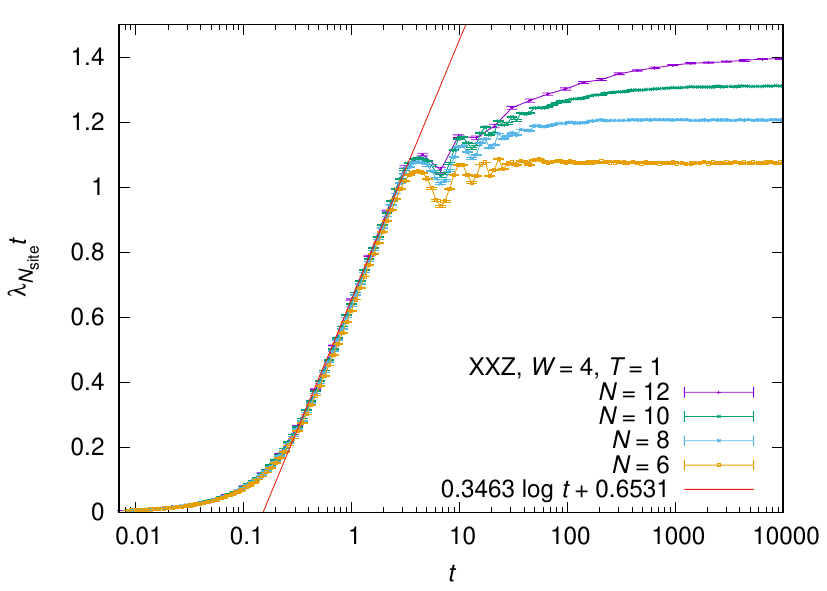}
\includegraphics[width=5cm]{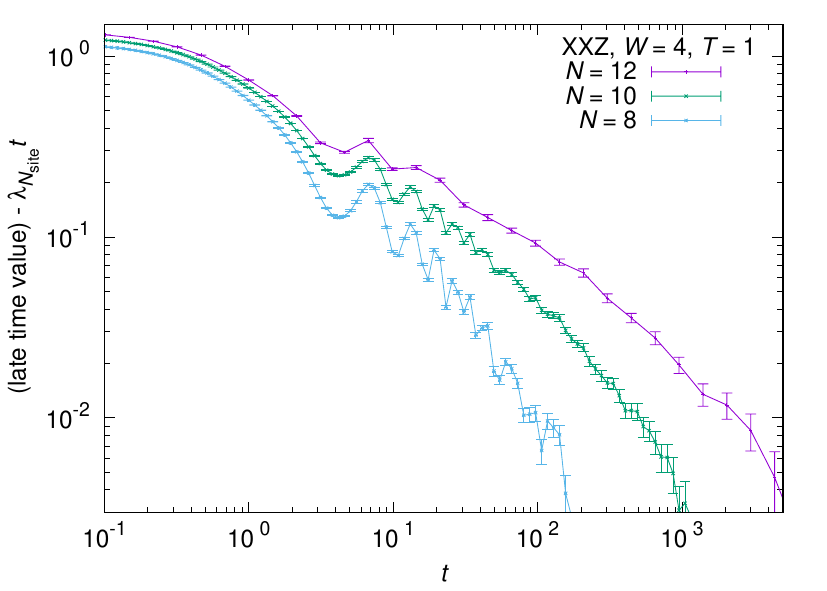}
\caption{
[Left] XXZ model, $\langle \lambda_{N_{\rm site}} t\rangle$, $W=0.5$ (ergodic phase), $T=1$.
The growth starts as $\lambda_{N_\mathrm{site}}t\sim 0.22 t$ (usual exponential growth),
then accelerated to $\lambda_{N_\mathrm{site}}t\sim 1.1 t^{1.5}$,
and then a power growth $\lambda_{N_\mathrm{site}}t\sim 0.36 \log t$ is observed all the way up to the plateau.
[Middle]
 $\langle \lambda_{N_{\rm site}} t\rangle$, $W=4.0$ (MBL phase), $T=1$.
The growth starts as $\lambda_{N_\mathrm{site}}t\sim 0.78 t$ (usual exponential growth),
then from $t\sim 0.3$ the power growth $\lambda_{N_\mathrm{site}}t\sim 0.35 \log t$ sets in,
but unlike the case of $W=0.5$, it ends at $t\sim O(N_{\rm site}^0)$ before the plateau is reached at exponetially long times.
[Right] The late-time behavior of $\langle \lambda_{N_{\rm site}} t\rangle$, $W=4.0$ (MBL phase), $T=1$.
The deviations from the late time values are plotted for several values of $N_{\rm site}$.
}
\label{fig:XXZ-lyap-growth_chaotic}
\end{figure}

In Fig.~\ref{fig:XXZ-lyap-growth_chaotic}, we plot $\lambda_{N_{\rm site}} t$ in order to see the detail of the Lyapunov growth.
The left figure is the ergodic phase, $W=0.5$.
The exponential growth $\lambda_{N_{\rm site}} t\sim t, \lambda_{N_{\rm site}}\sim 0.3$ can be seen at early time.
At some intermediate $O(N_{\rm site}^0)$ time, the power-law growth sets in.
Similar behaviors both in ergodic and MBL phases ($W=4.0$) shown in the middle.
However the late-time behaviors are rather different. In the ergodic phase, the power growth continues
all the way up to the plateau, which scales $\sim\log N$.
On the other hand, in the MBL phase, the power growth stops at $O(N_{\rm site}^0)$ time,
and much slower growth sets in.
In the left panel of Fig.~\ref{fig:XXZ-lyap-growth_chaotic}, the deviation from the late-time plateau in the MBL phase is plotted
in the log-log scale. We can see a straight line, which means
the late-time behavior is $A-Bt^{-p}$.
This is consistent with the theoretical expectation in Ref.~\cite{Swingle:2016jdj}.
With the range of $N_{\rm site}$ available at this moment, it is hard to take the large volume limit, $N_{\rm site}\to\infty$.

A possible explanation of this pattern is as follows. In the classical theory,
the perturbation at $t=0$ can be sent arbitrarily small, and the exponential growth can continue forever.
However when the perturbation is finite, the exponential growth stops at a finite time;
otherwise the causality is broken! (In the nonrelativistic theory the speed is not limited,
but once the initial condition is set, then the energy conservation sets the upper limit of the speed at later time.)

In the quantum theory, the perturbation is necessarily finite and hence the exponential growth has to stop at some point.
In nonlocal systems like the matrix model and SYK model, the exponential growth stops when
the `local' perturbation (say the multiplication of $\psi_1$) affects all other degrees of freedom substantially.
This is typically $O(\log N_\mathrm{site})$ time.
For local systems like the spin chain, the exponential growth has to stop at $O(1)$ time, because $\sigma_j(t)$
and $\sigma_i(0)$ commute when $t$ is smaller than $|i-j|$ divided by the butterfly velocity;
when the exponential growth stops, only $O(1)$ number of degrees of freedom talk to each other.
(The system size does not matter, otherwise the causality or the Lieb-Robinson bound is broken.)
$\hat{M}_{ij}(t)$ is a banded matrix with width $w\sim t$, and if we use a very rough approximation that
all the nonzero entries are of order one, then the singular values scale as $\sqrt{w}$.
It leads to a late time behavior $\lambda t\sim 0.5\log t$, which is in the right ballpark compared to $0.35\log t$
and $0.32\log t$ in the left panels of Fig.~\ref{fig:XXZ-lyap-growth_chaotic}.

As a related example, let us consider planar black $p$-brane ($p>0$), which is described by U$(N)$ super Yang-Mills on ${\mathbb R}^{1,p}$.
How is a localized perturbation scrambled in this theory? Firstly the fast scrambling with $\lambda\sim 2\pi T$
mixes the information among the gauge degrees of freedom; then the information gradually spreads along ${\mathbb R}^{p}$.
In terms of gravity, the horizon has a topology of S$^{8-p}\times{\mathbb R}^p$, and the fast scrambling takes place along S$^{8-p}$
while the growth along ${\mathbb R}^p$ is slower and dominant at late time.
The 1d spin chain is analogous to $p=1$ and very small $N$.

In the explanation above, only the local physics is important for the early-time exponential growth.
Hence the same pattern is expected both in the ergodic and MBL phases.
The time scale of the saturation of the power growth can be different; in the ergodic phase the saturation time scale should increase
with the system size, while in the MBL phase it is independent of the system size,
instead only the volume of the region affected by the perturbation matters.
It is consistent with the numerical results:
in Fig.~\ref{fig:XXZ-lyap-growth_chaotic} the saturation time scale increases with the system size,
while in Fig.~\ref{fig:XXZ-lyap-growth_chaotic} it seems to be insensitive to the system size.

For local quantum systems, the absence of the exponential Lyapunov growth should be generic.
As we have seen above, it is not easy to distinguish the ergodic and MBL phases just from the power growth.
However, as we will see in Sec.~\ref{sec:XXZ-RMT}, the statistical features of the Lyapunov exponents are clearly different
in these two phases.

\begin{figure}
\includegraphics[width=16.5cm]{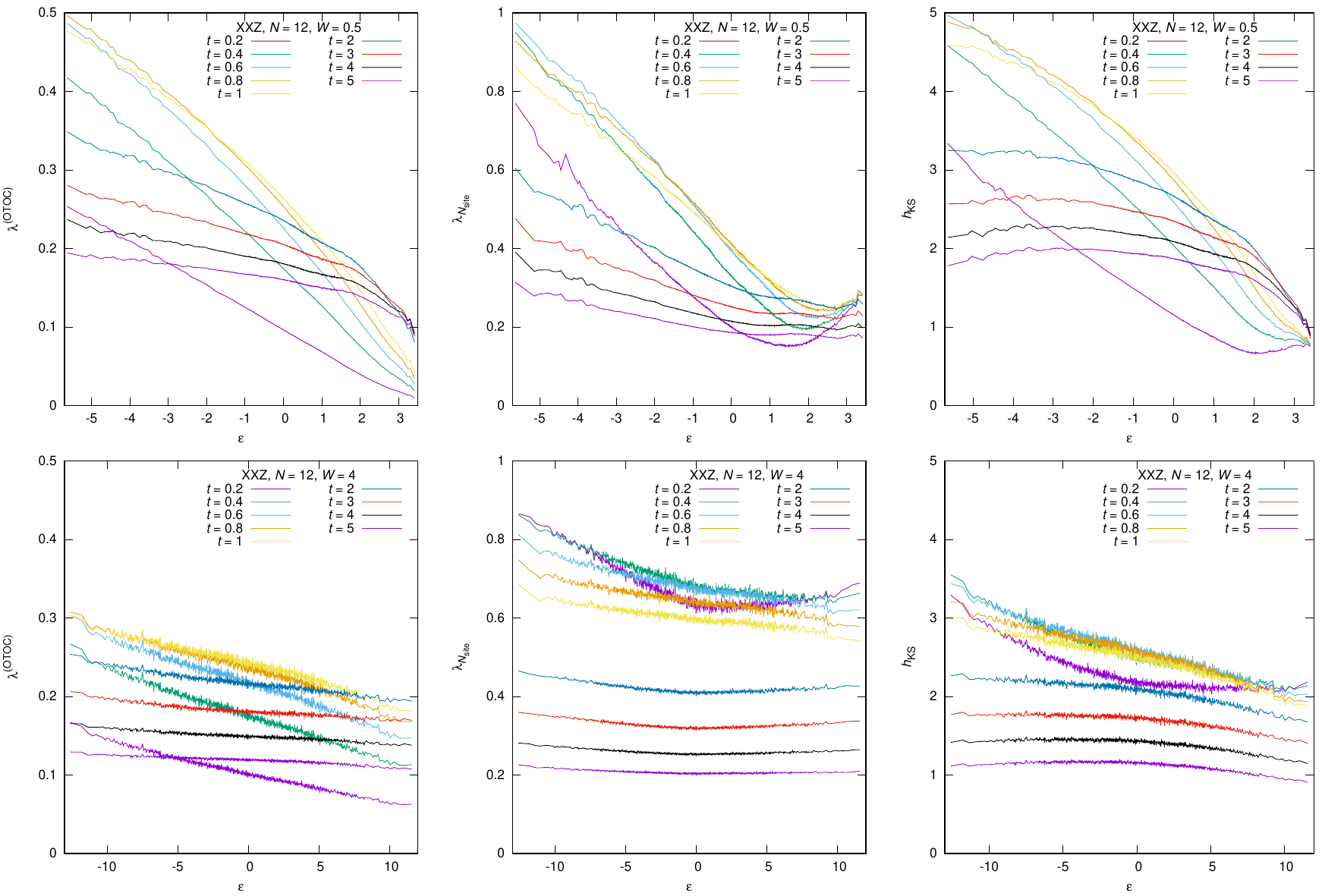}
\caption{
For XXZ model, the dependence on the eigenstate energy of $\lambda_\mathrm{OTOC}$, $\lambda_N$ and $h_\mathrm{KS}$ are plotted.
For $i=1,2,\ldots,924$, the sample average of these quantities for the eigenstate with the $i$-th smallest energy eigenvalue of the Hamiltonian is plotted against the average of the energy.
Upper: $W=0.5$. Lower: $W=4$.}
\end{figure}

\section{Random Matrix Statistics of Lyapunov Spectrum}
\hspace{0.51cm}

In this section we study the statistical properties of the quantum Lyapunov spectrum,
motivated by the universality in the classical Lyapunov exponents explained in Sec.~\ref{sec:universality_classical}.

\subsection{Lyapunov Spectrum vs RMT in SYK}\label{sec:SYK-RMT}
\hspace{0.51cm}
According to Ref.~\cite{Garcia-Garcia:2017bkg,Nosaka:2018iat},
the $q=2$ deformed SYK model (Eq.~\eqref{eqn:q=2deformedSYK}) is integrable at sufficiently low temperature, when $K>0$.
Therefore, by carefully choosing the energy eigenstates, we can study the statistical features of the Lyapunov spectra
in the integrable and chaotic phases, in principle.
However, as we have seen already, perturbations by multiplications of $\hat{\psi}$ is not really `small'
at the system size we can study numerically, and hence it is hard to study the properties of the integrable phase
and chaotic phase separately.

Below, in addition to the nearest-neighbor level correlation,
we study the nearest-neighbor gap ratio,
\begin{eqnarray}
r_i
=
\frac{ \min(\lambda_{i}-\lambda_{i-1},\lambda_{i+1}-\lambda_{i})}{\max(\lambda_{i}-\lambda_{i-1},\lambda_{i+1}-\lambda_{i})}.
\end{eqnarray}

\begin{figure}[htbp]
\begin{center}
\includegraphics[width=16cm]{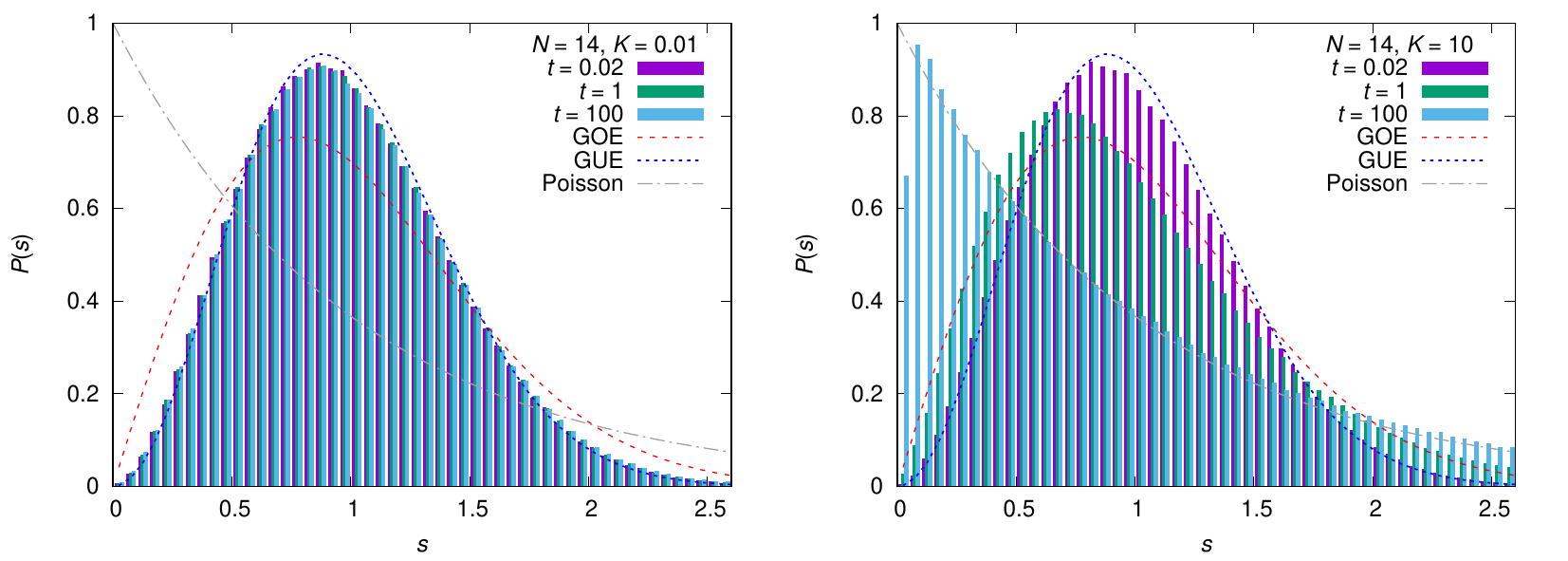}
\includegraphics[width=16cm]{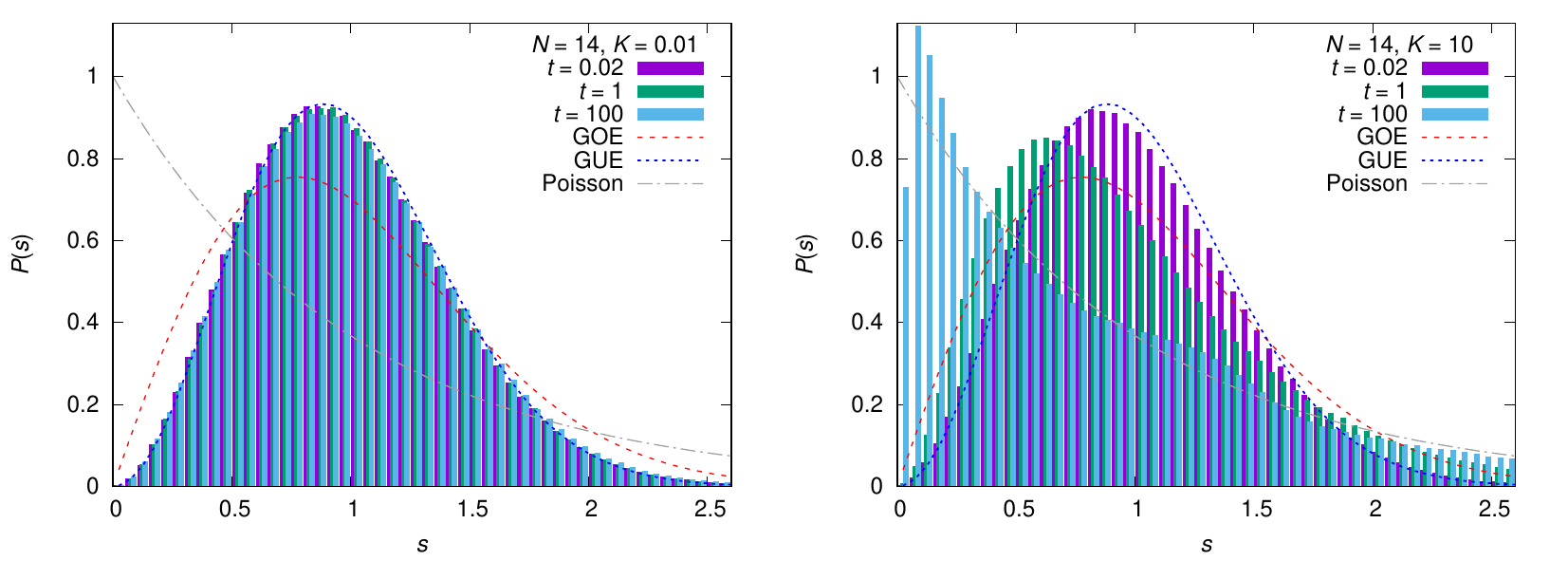}
\end{center}
  \caption{
Nearest neighbor correlation for the unfolded Lyapunov spectrum of the generalized Majorana SYK model,
$N=14$, $\beta=0$, $0.02\leq t \leq 100$ with $K=0.01, 10$.
[Upper] standard unfolding is conducted by fitting the density of exponents using a polynomial.
[Lower] The fixed-$i$ unfolding has been performed.
When $K$ is small, good agreement with GUE is observed, until very late time.
When $K$ is large, small deviation from GUE can be seen at early time, and the deviation grows quickly at later time.
}\label{fig:N14upper-standard-and-fixed-i-unfolding}
\end{figure}

In the upper row of Fig.~\ref{fig:N14upper-standard-and-fixed-i-unfolding}, we have shown the nearest neighbor level separation
estimated after the standard unfolding procedure.
Namely, we have estimated the distribution of the Lyapunov exponents by using energy eigenstates within certain range
(between 5\% and 10\% in these specific plots),
numerical fit it by polynomial of degree 10 and used it for the unfolding.
We can see good agreement with GUE at small $K$, but there is a small but visible deviation.

A possible flaw of this method is that, when $N_\mathrm{site}$ is small, peaks arising due to the level repulsion can be visible
and the unfolding can eliminate them as well, so that the universal random matrix behavior is erased.
To circumvent such possibility, we tried another unfolding prescription as well:
normalize the gap $g_i=\lambda_{i+1}-\lambda_i$ so the average is 1. Define  $\tilde{g}_i\equiv g_i/\langle g_i\rangle$
and look at the distribution of $\tilde{g}_i$.
We call it `fixed-$i$ unfolding'.
The result is shown in the lower row of Fig.~\ref{fig:N14upper-standard-and-fixed-i-unfolding}.
Compared to the standard unfolding (the upper row of Fig.~\ref{fig:N14upper-standard-and-fixed-i-unfolding}),
the agreement with GUE is improved substantially for $K=0.01$. On the other hand, for $K=10$, there is no improvement.

\begin{figure}[htbp]
\begin{center}
\includegraphics{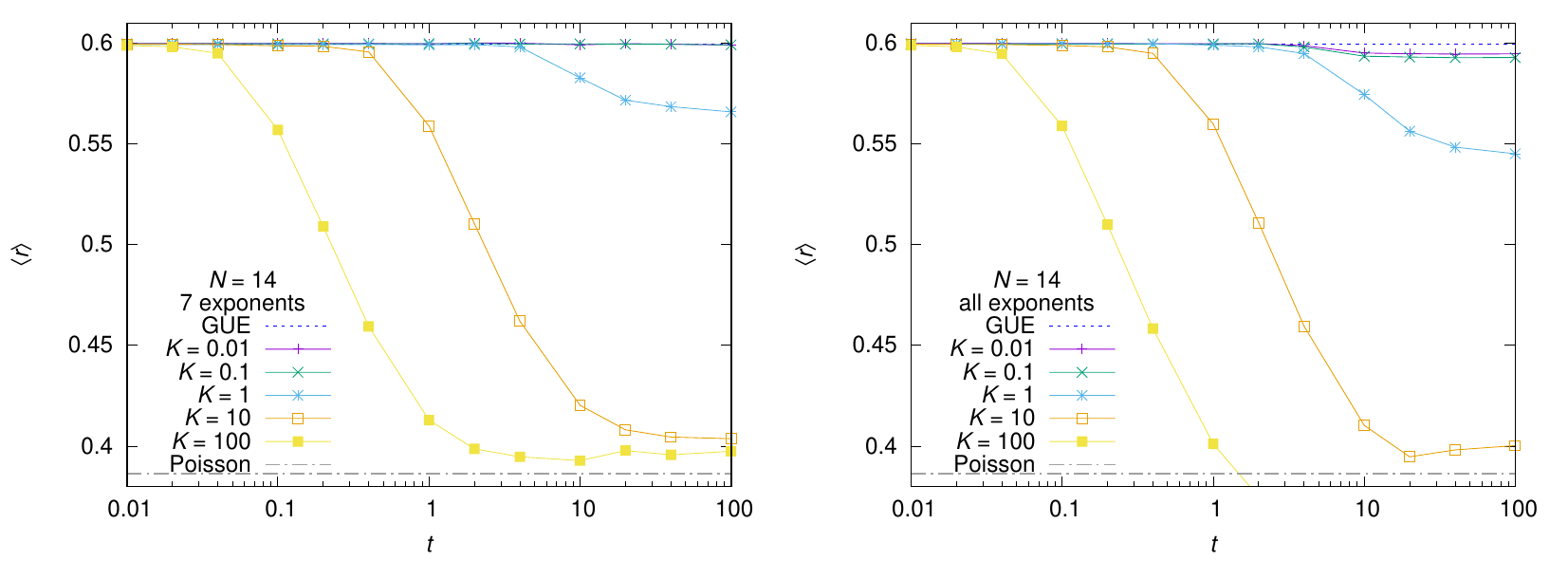}
\end{center}
\caption{
The $r$-parameter $\langle r\rangle$ for the $N=14$ SYK model, with fixed-$i$ unfolding, obtained by
[left] using the larger $N/2 = 7$ exponents, and [right] using all the exponents.
The early rime behaviors are almost identical.
(Different late time behaviors appear because a large gap sets in between $\lambda_{N/2}$ and $\lambda_{N/2+1}$.)
At $K=10$ and $100$, large deviation from the GUE value is observed before
the Lyapunov growth (with almost constant Lyapunov exponent) sets in.
At $K=1$, a large deviation is observed before the Lyapunov growth ends.
At $K=0.01$ and $0.1$, the $r$-parameter stays close to the GUE value
even at $t=100$.
}
\label{fig:SYK-r-parameter}
\end{figure}

\begin{figure}[htbp]
\includegraphics[width=16cm]{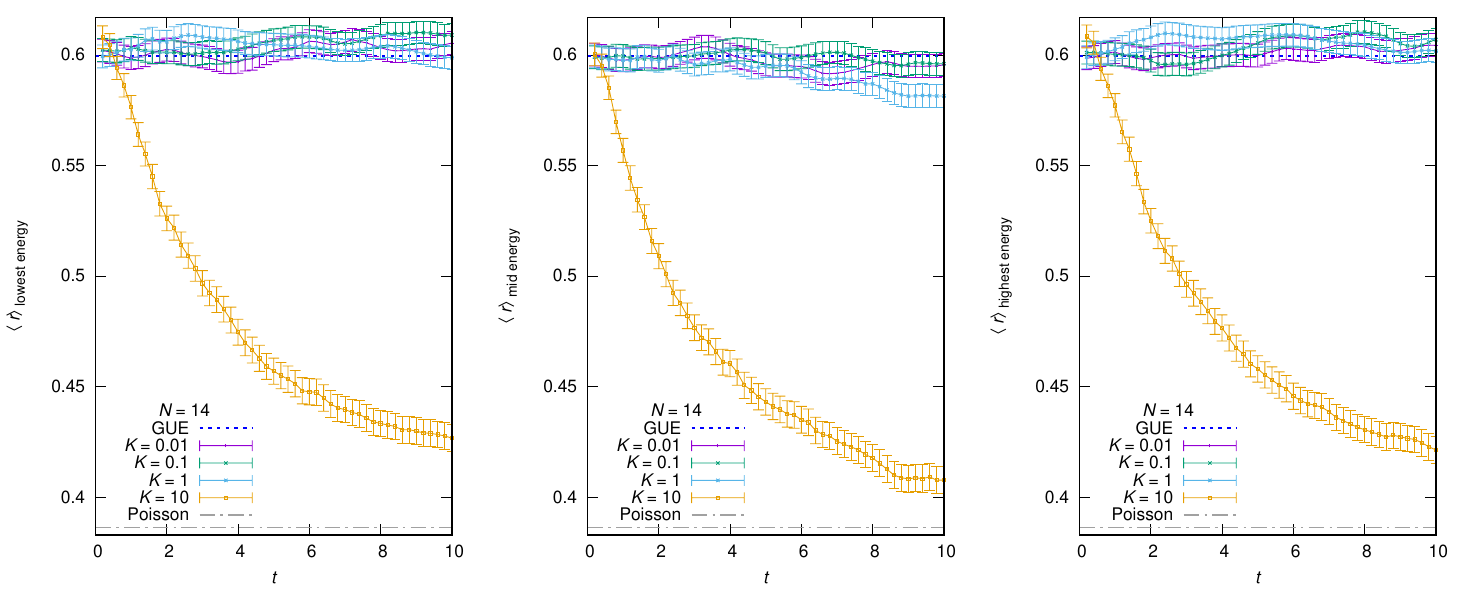}
\caption{
The $r$-parameter $\langle r\rangle$ with fixed-$i$ unfolding, obtained using all exponents for the $N_\mathrm{site}=14$ SYK model, for the ground state (left), the state at the center of the energy spectrum (center) and the highest energy state, plotted as a function of $t$ for $K=0.01, 0.1, 1, 10$.
}
\label{fig:SYK-r-N14all}
\end{figure}

In Fig.~\ref{fig:SYK-r-parameter}, we show the time dependence of $r$ at $\beta=0$.
(As discussed around Fig.~\ref{fig:SYK-Lyapunov-spectrum-beta=0} for the case of $N=16$,
a gap develops between $\lambda_{N/2}$ and $\lambda_{N/2+1}$ for larger $K$ as $t$ is increased.
In order to check that this gap does not affect the result, we have also calculated $\langle r\rangle$ using only the larger $N/2$ exponents.
We can see that the early rime behaviors are almost identical.)
At $K=10$ and $100$, large deviation from the GUE value is observed before
the Lyapunov growth (with almost constant Lyapunov exponent) sets in.
At $K=1$, a large deviation is observed before the Lyapunov growth ends.
At $K=0.01$ and $0.1$, the $r$-parameter stays close to the GUE value
even at $t=100$.
In Fig.~\ref{fig:SYK-r-N14all}, we show essentially the same plot, but using $|E_0\rangle$, $|E_{L/2}\rangle$
and $|E_{L-1}\rangle$. We don't see a significant change as expected; for the value of $N$ we study, the perturbation is too large,
so that we can only see a mixture of almost all states.
At sufficiently large $N$ we expect different behaviors depending on the energy.

As $K$ becomes larger, more energy eigenstates belong to the integrable sector.
That the Poisson statistics sets in with larger $K$ suggests that the spectrum in the integrable sector follows Poisson.
Note that the GUE appears as $t\to 0$ even at large $K$. We do not have understanding about this property.

\subsection{Lyapunov Spectrum vs RMT in XXZ}\label{sec:XXZ-RMT}
\hspace{0.51cm}
Because the total $z$-spin $S^{\rm (total)}_z$ commutes with the Hamiltonian \eqref{eqn:XXZ},
we focus on the Lyapunov spectra obtained by using eigenstates with $\langle S^\mathrm{(total)}_z \rangle = 0$.

As explained in Sec.~\ref{sec:XXZ_definition}, this theory has ergodic and MBL phases.
We will study $W=0.5$ and $1.0$, which are mainly in the ergodic phase except for the edges of the energy spectrum,
and $W=4.0$, which is dominantly in the MBL phase.

\begin{figure}[htbp]
\centering
\includegraphics[width=15cm]{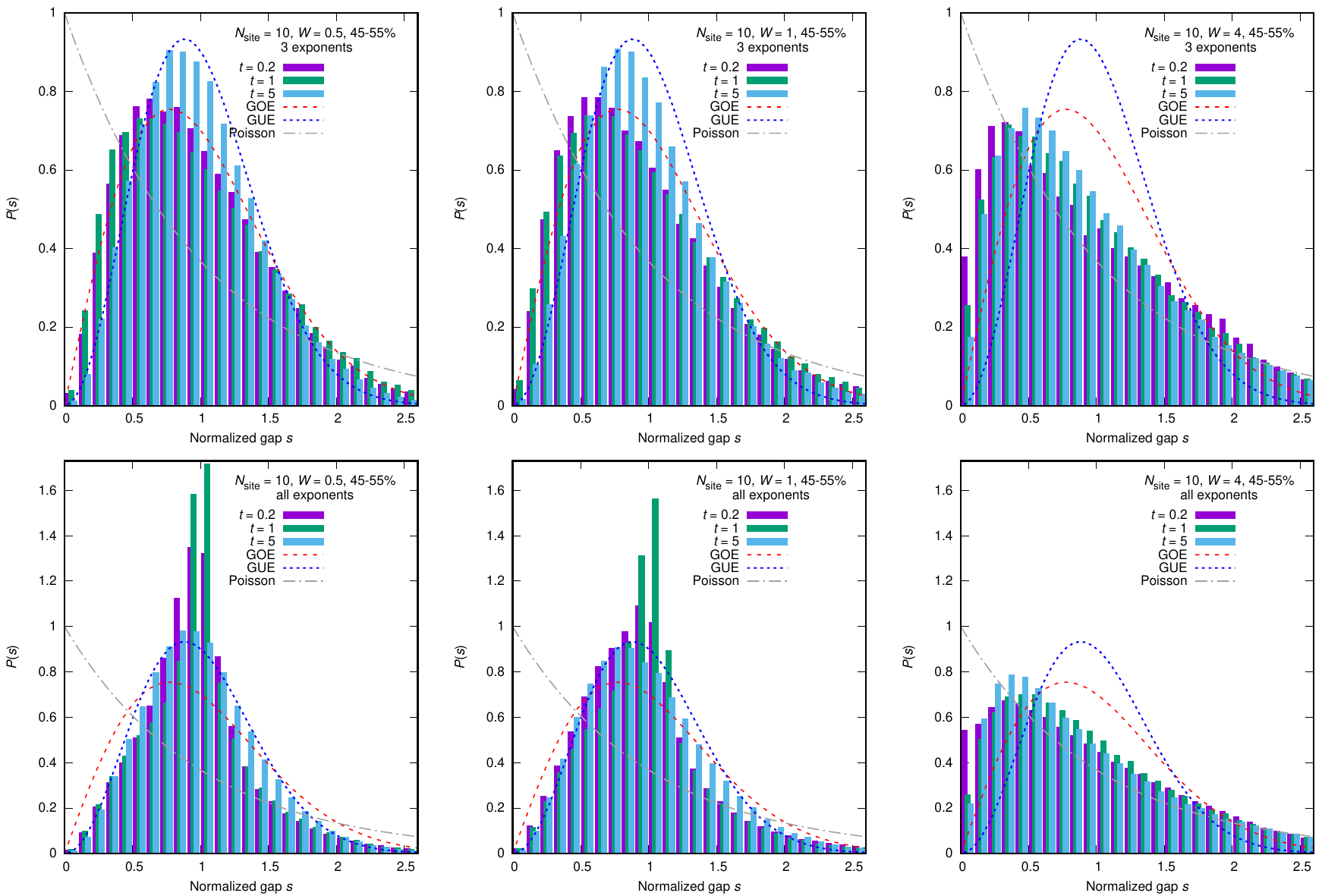}
\caption{
XXZ model, distribution of the normalized gap $P(s)$.
$N_\mathrm{site}=10$ and $900$ samples have been used for each case.
The fixed-$i$ unfolding [upper] for the two gaps between the three largest exponents and [lower] for all gaps have been used for the Lyapunov exponents obtained using 45\% -- 55\% of the spectrum for each sample, in which 0\% corresponds to the ground state.
Left: $W=0.5$. Middle: $W=1$. Right: $W=4$.
}
\label{fig:PsXXZ5-10p}
\end{figure}

\begin{figure}[htbp]
\centering
\includegraphics[width=15cm]{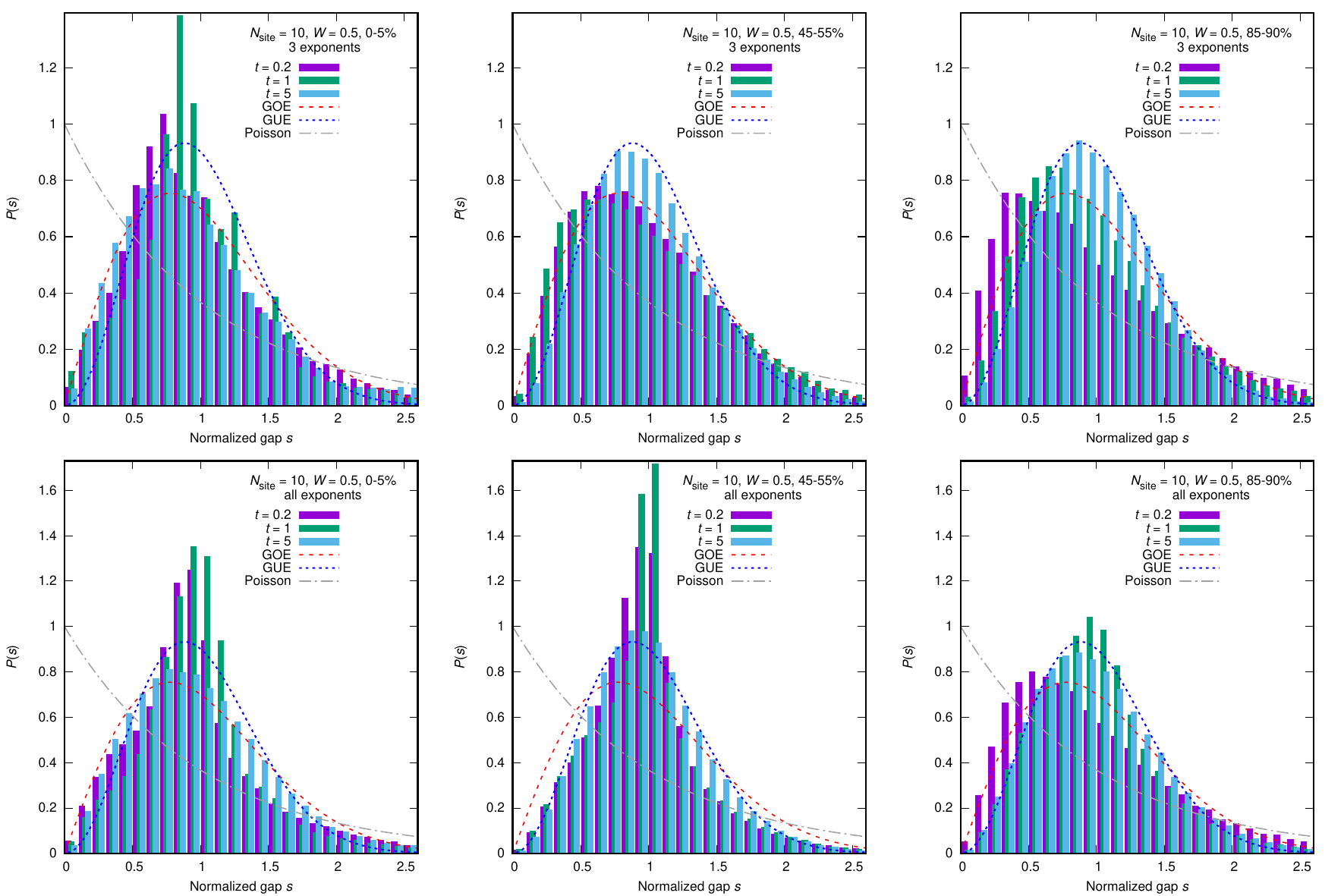}
\caption{
XXZ model, distribution of the normalized gap $P(s)$.
$N_\mathrm{site}=10$, $W=0.5$ and $900$ samples have been used for each case.
The fixed-$i$ unfolding [upper] for the two gaps between the three largest exponents and [lower] for all gaps have been used for the Lyapunov exponents obtained using 0\% -- 5\% (left), 45\% -- 55\% (middle;
the same as the left panel in Fig.~\ref{fig:PsXXZ5-10p}),
and 85\% -- 90\% (right) of the spectrum for each sample.
}
\label{fig:PsXXZ-various-range}
\end{figure}

In Fig.~\ref{fig:PsXXZ5-10p} we have plotted $P(s)$ obtained by using eigenstates with the energy within
45\% -- 55\% from the lower edge of the spectrum,\footnote{
We have used such a narrow energy band because we can actually see the energy dependence unlike the case of SYK,
as explained in Sec.~\ref{sec:perturbation_size}.}
with the fixed-$i$ unfolding introduced in Sec.~\ref{sec:SYK-RMT}.
We can see a good agreement with GUE for $W=0.5$ and $1.0$ (the ergodic phase) at sufficiently late time,
while the Poisson distribution is favored for $W=4.0$ (the MBL phase).
In Fig.~\ref{fig:PsXXZ-various-range}, we fixed $W=0.5$ and varied the energy band.
We can see the time evolution strongly depends on the choice of the energy.
We can also see that the GUE is not obtained near the ground state, which is close to the MBL phase.
Note that we have shown two results, one obtained by using all exponents and the other obtained by only the largest three exponents.
The reason is as follows.
When we plot the sample averaged values of $\lambda_i$ against the eigenstate index, for the XXZ model with smaller values of $W$, at short times large gaps between the smaller, nearly twofold degenerate exponents are observed for lower energy eigenstates. For larger $N/2$ exponents, the averaged values are evenly distributed for $t\gtrsim 5$.
Therefore, in order to make sure the universal behavior can be seen regardless of the choice of the exponents, we have shown two results.
Below, we will also study the gap ratio $r$. This is more sensitive to the change of the gap size, and hence, we will use the largest three exponents for safety.

\begin{figure}[htbp]
\centering
\includegraphics[width=15cm]{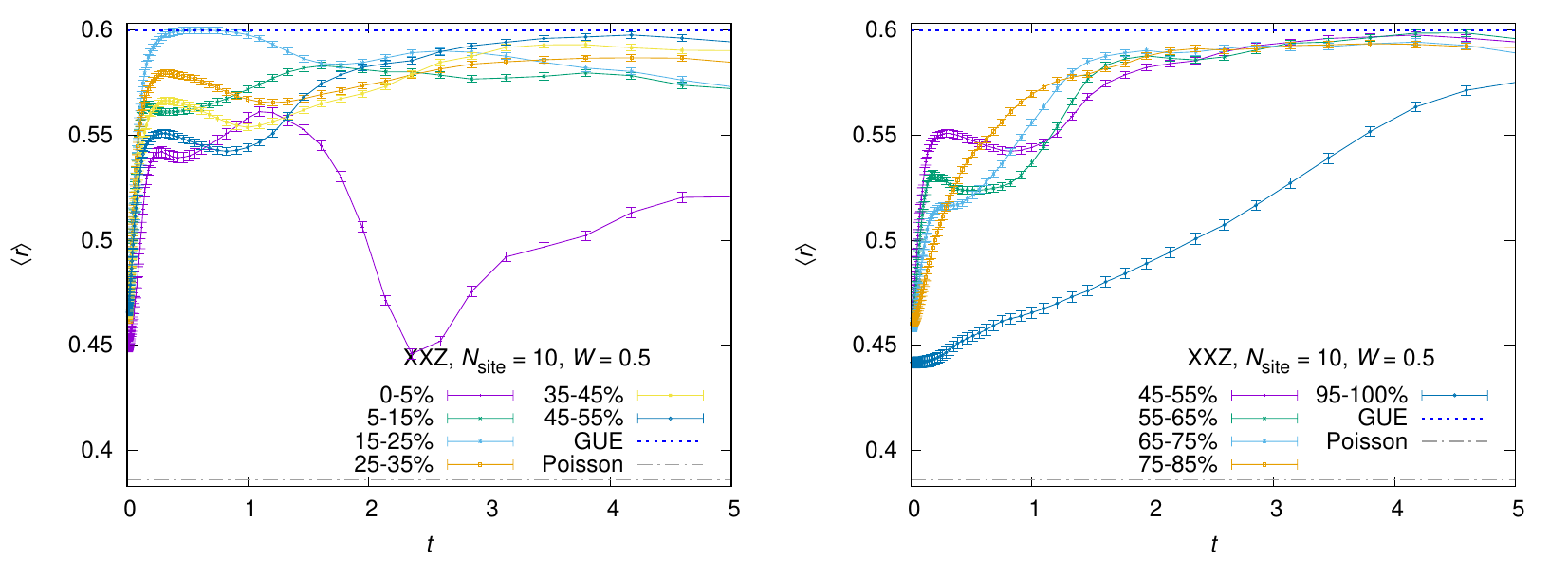}
\caption{
Plot of $\langle r\rangle$ for the two gaps between the three largest exponents for XXZ, $N_\mathrm{site}=10$, $W=0.5$. Better agreement with GUE (both the value and time window) can be seen at the center of the energy spectrum,
which is consistent with the fact that the center of the energy spectrum is in the ergodic phase while the edges remain MBL.
}\label{r-XXZ-W05}
\end{figure}

In order to see the time and energy dependence quantitatively, we have plotted $\langle r\rangle$ with the fixed-$i$ unfolding
in Fig.~\ref{r-XXZ-W05}, for $W=0.5$, for various energy bands.
We can see better agreement with GUE (both the value and time window) at the center of the energy spectrum.
Recalling that the middle of the energy spectrum is in the ergodic phase except that the small region near the edges remain MBL,
this is consistent with the interpretation that GUE is obtained for the ergodic phase but not for the MBL phase.

\begin{figure}[htbp]
\centering
\includegraphics[width=15cm]{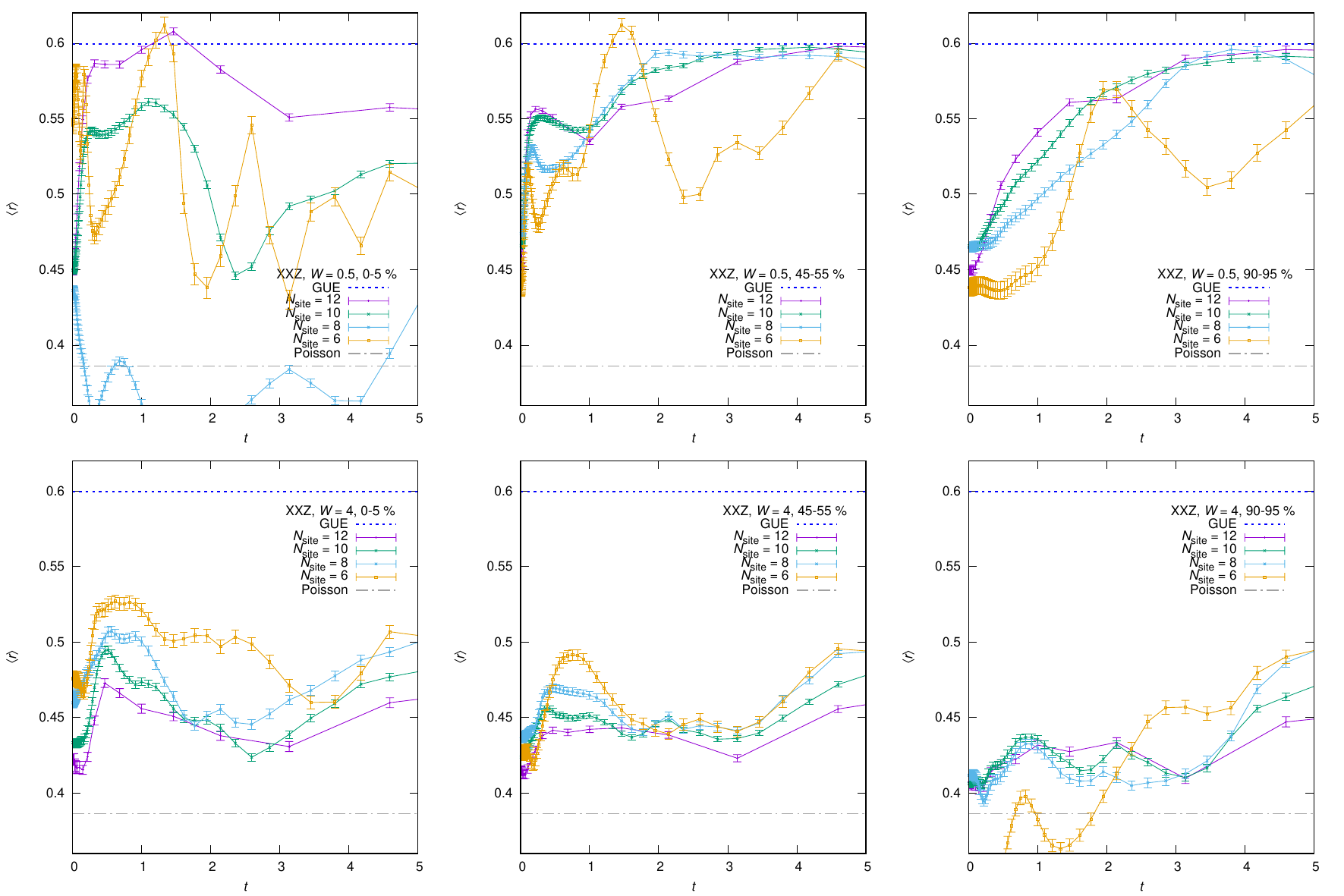}
\caption{
XXZ model, plot of averaged nearest gap ratio $\langle r\rangle$, obtained using individual gap unfolding for the two gaps between the three largest exponents for $N_\mathrm{site}=12, 10, 8, 6$.
Top: $W=0.5$. Bottom: $W=4$. Both in upper and lower rows, the
plots are obtained using the eigenstates in between [left] $0\%$ and $5\%$, [middle] $45\%$ and $55\%$, and [right] $90\%$ and $95\%$ of the energy spectrum for each sample.
}\label{r-XXZ-N-dependence}
\end{figure}

The $N_{\rm site}$-dependence of $\langle r\rangle$ is shown in Fig.~\ref{r-XXZ-N-dependence}. The GUE can be seen with good precision at $N_{\rm site}\ge 8$.
We studied the values of $\langle r\rangle$ for $N=6,8,10$ and $12$ until very late time ($t\lesssim 10^8$).
For these values of $N$, $\langle r\rangle$ becomes almost constant, which is different from the GUE value,
at $t\gtrsim 10$. Therefore we expect the importance of the large-$N$ limit before $t\to\infty$ for the emergence of the universality.

\section{Conclusion and Outlook}
\hspace{0.51cm}
In this paper we have proposed a generalization of the Lyapunov spectrum to quantum theories,
and studied its properties by using the SYK model and the XXZ model with random magnetic field as examples.
By definition, the Lyapunov spectrum contains more information than just the largest exponent.

The KS entropy --- which we defined by the sum of the positive exponents --- is likely to be a better characterization of the strength of the chaos,
because it can describe the entropy production rate.
We conjectured that the black hole maximized the KS entropy.

We also found the numerical evidence for the universality of the Lyapunov spectrum.
(Previously, this universality has been observed in classical chaos as well \cite{Hanada:2017xrv}.)
It is interesting if we could understand the meaning of the onset of the universal RMT behavior.
It should have something to do with holography, because special theories which are dual to quantum black holes
--- the SYK model, and classical D0-brane matrix model, as demonstrated in Ref.~\cite{Hanada:2017xrv} --- show the universality already at $t=0$.
We propose that the quantum systems holographically dual to Einstein gravity satisfy this `strong' universality.

The conjectures above are based mainly on the numerical observations for limited number of theories.
It is important to study more examples, and also, to develop the understanding on the gravity side.
Another important issue is how the universality class is determined. For the examples studied in this paper
we observed only GUE ensemble, regardless of the system size.

It is also important to apply the method presented in this paper to various physical systems,
especially in the contexts of condensed matter and quantum gravity.
It should be possible to get new insight into scrambling and thermalization by observing the Lyapunov growth,
and with various examples we might be able to understand the meaning of the characteristic time scales
associated with the Lyapunov growth and the onset of the universal spectral behavior.
Toward the study of full string theory, probably the weakly coupled region of D0-brane matrix model \cite{BHS-in-progress,Buividovich:2017kfk}
is a good place to start.

It would also be interesting to develop a measurement protocol for the Lyapunov spectrum along the lines on Ref.~\cite{swingle2016}. A brute force way to approach the problem is to consider performing a whole set of many-body interference experiments to measure the various matrix elements needed to construct the spectrum-defining matrix. A detailed study of the feasibility of the this approach, and the search for more economical approaches, is left to future work.

\section*{Acknowledgements}
\hspace{0.51cm}
We thank P.~Buividovich, A.~M.~Garc\'ia-Garc\'ia, G.~Gur-Ari, B.~Loureiro, C.-T.~Ma, M.~Nozaki,
P.~Romatschke, A.~Romero-Berm\'udez, S.~Shenker,
H.~Shimada and J.~Yoon for useful discussions.
This research was supported in part by the International Centre for Theoretical Sciences (ICTS) during a visit (M. H.) for the program ``Nonperturbative and Numerical Approaches to Quantum Gravity,  String Theory and Holography",
where a part of the results in this paper have been presented. H.G. was supported in part by NSF grant PHY-1720397.
M.~H. and M.~T. thank the Yukawa Institute for Theoretical Physics at Kyoto University; the discussions during the YITP-T-18-04 ``New Frontiers in String Theory 2018'' were useful to complete this work.
We acknowledge JSPS KAKENHI Grants 17K14285 (M.~H.) and 17K17822 (M.~T.). This work was also partially supported by the Simons Foundation via the It From Qubit Collaboration, and by the Department of Energy, award number DE-SC0017905.

\bibliographystyle{utphys}
\bibliography{universality}

\end{document}